\documentclass[%
 reprint,
 amsmath,amssymb,
 aps,
]{revtex4-1}

\usepackage{graphicx}% Include figure files
\usepackage{dcolumn}% Align table columns on decimal point
\usepackage{bm}% bold math

\usepackage{grffile}
\usepackage{float}

\usepackage{amsmath}
\usepackage{marvosym}
\usepackage{amsfonts}
\usepackage{xcolor}
\usepackage{amssymb}
\usepackage{morefloats}
\usepackage{caption}

\newcommand{\threefigl}[6][0]{
\begin{figure}[H]
\centering
  \includegraphics[width=#4\hsize]{#1}
  \includegraphics[width=#4\hsize]{#2}
  \includegraphics[width=#4\hsize]{#3}
  \caption{#5}
\label{#6}
  \nobreak%\medskip
\end{figure}}

\newcommand{\twofigl}[5][0]{
\begin{center}
  \vspace{.1cm}
  \includegraphics[width=#3\hsize]{#1}
  \includegraphics[width=#3\hsize]{#2}
   \captionof{figure}{#4}
\label{#5}
  \nobreak%\medskip
\end{center}}

\begin{document}

\preprint{APS/123-QED}

%\title{An improved scheme for detection of gravitational waves using CNN and TDA}% Force line breaks with \\
%\title{Detection of gravitational waves using topological data analysis and convolutional neural network: An improved neural network approach}
\title{Detection of gravitational waves using topological data analysis and convolutional neural network: An improved approach}

\author{Christopher Bresten}
\email{cbresten@ajou.ac.kr}
 \affiliation{Department of AI and Data Science, Ajou University, Suwon 16499, Korea}%Lines break automatically or can be forced 

%with \\
\author{Jae-Hun Jung}%
 \email{jaehunjung@ajou.ac.kr, jaehun@buffalo.edu}
% \altaffiliation[Also at ]{}%Lines break automatically or
 \affiliation{Department of AI and Data Science, Ajou University, Suwon 16499, Korea \\ \& Department of Mathematics, University at Buffalo, State University of New York, Buffalo, NY 14260-2900, U.S.A.}%

%\collaboration{UMass Dartmouth}%\noaffiliation

%\affiliation{
%\affiliation{

\date{\today}

\begin{abstract}
The gravitational wave detection problem is challenging because the noise is typically overwhelming. Convolutional neural networks (CNNs) have been successfully applied, but require a large training set and the accuracy suffers significantly in the case of low SNR. We propose an improved method that employs a feature extraction step using persistent homology. The resulting method is more resilient to noise, more capable of detecting signals with varied signatures and requires less training. This is a powerful improvement as the detection problem can be computationally intense and is concerned with a relatively large class of wave signatures.

\end{abstract}

%Use showkeys class option if keyword
\maketitle

%\tableofcontents

%\section{\label{sec:level1} Background and motivation}

\section{Background and motivation}
The pioneering work by Huerta et al. \cite{huerta} showed convolutional neural networks (CNN) to be a powerful method approach to the gravitational wave (GW) detection problem -- a GW signature buried in the noisy interferometer data \cite{ligo0,ligo1} can be detected with a CNN.  A CNN is a regularized multilayer artificial neural network that utilizes the hierarchical features in data \cite{cnn0,cnn1}.  The localized nature of convolutions makes CNNs demonstrate great performance on raw data, especially image data.   The approach is to train the CNN with noisy data with and without GW signatures. Data streams from interferometers are searched by matched-filter with a collection of approximate GW templates. When high correlation with a template is detected, it is shared with partners for verification with data sources from other interferometers and electromagnetic follow-up \cite{ligo1}. Appropriately trained CNNs have shown to be an excellent first pass detection method to proceed the more specific computationally expensive matched-filter and labor intense verification steps.

The original method of \cite{huerta} followed the standard CNN architecture, so requires a large training set and a computationally cumbersome choice of hyper-parameters. Adding more feature selection prior to classification with CNN can improve performance. %Furthermore, the network architecture, not unique by nature, needs to be tuned for better performance. Adding more intelligent features  prior to classification with the CNN can overcome some of these difficulties. 

%In this letter, we proposed and tested a new method of improving the CNN approach by considering the topology of the GW signal and concatenating it with the CNN, particularly by measuring homology of signals. 

In this letter, we propose a new method improving on the CNN approach by including topological features of the data, in particular,  persistent homology of sliding window embeddings. This is known as topological data analysis (TDA) \cite{topologyanddata}.  The proposed method makes training more efficient, consequently reducing the size of the training set significantly. The aforementioned localized effect of convolution layers makes this a low-risk endeavor, as adding topological features should not decrease performance because as the CNN is trained, it can ignore these features by assigning small weights.

%This was done by adding the desired homology of GW signals  in the sliding window embedding space.  The aforementioned localized effect of convolution layers makes this a low-risk endeavor and adding topological features does not decrease performance because CNN is trained as if it ignores these features with small weights during training. 

The important potential enhancement is the increased generality. Interesting GW signals  come in a large and diverse class. For instance,  multiple parameters  are involved in a black-hole merger that change the signature, e.g. the mass ratio. TDA is a lossy process, but preserves  various key  properties such as period, decay rate, etc. when classifying wave-packets.

\section{Data synthesis}\label{sec:datasyn}
Signals were generated by a surrogate model described in \cite{blackmodel}. The model generates non-spinning binary black-hole merger gravitational waveforms with mass ratio between $1$ and $10$. It has an accuracy close to that of the high-fidelity model which requires solving Einstein's equations by the {\it Spectral Einstein Code} (SpEC). The reduced model is constructed by selecting most relevant mass ratios using a greedy algorithm. The surrogate model is highly accurate after including about $15$ waveforms, in that the error becomes comparable to the truncation error of the SpEC.

We use $1500$ reference signals with mass ratios between $1.0078$ and $9.9759$, sampled at $2048Hz$. Each window length is $2$ seconds. We construct training sets by adding noise and embedding the signal in noise so it occurs at a random time. 

Let $g$ be a signal from the reference set  and $\xi$ be the Gaussian noise with standard deviation $1$. The GW signal $g$ is embedded in $\xi$.  The non-Gaussian noise can be treated in the similar manner.  We scale the noise against the signal with a unitless scaling coefficient $R$. The synthetic data is then:
\begin{eqnarray}
s = g + \epsilon \frac{1}{R} \xi . 
\end{eqnarray}
The coefficient $\epsilon = 10^{-19}$ scales the noise amplitude down to roughly the same order of magnitude as the signal.  The signal is inserted at a random position in a piece of noise of duration $0.976$ seconds ($2000$ elements at $2048$Hz) scaled with the same factor as above. This yields a signal of length $\approx 2.976$ seconds. A GW signal is present with probability of $p = 0.5$, implying that the training data is balanced.  By cycling through a sample of signals and values for $0.075<R<0.65$ while randomly choosing signal presence and occurrence time, we construct an arbitrarily large training set.  The coefficient $R$ corresponds to the optimal match-filtered SNR in Table \ref{tab:snr}. 

\begin{table}[b]
\begin{ruledtabular}
\caption{\label{tab:snr} Sample $R$ and  corresponding SNRs. }
\begin{tabular}{c | c c c c c c}
$R$ & 0.075 & 0.19& 0.305& 0.42& 0.535 & 0.65\\ \hline
SNR & 2.097 &  5.327 & 8.523 & 11.56 & 14.79 & 17.98\\
\end{tabular}
\end{ruledtabular}
\end{table}

\threefigl[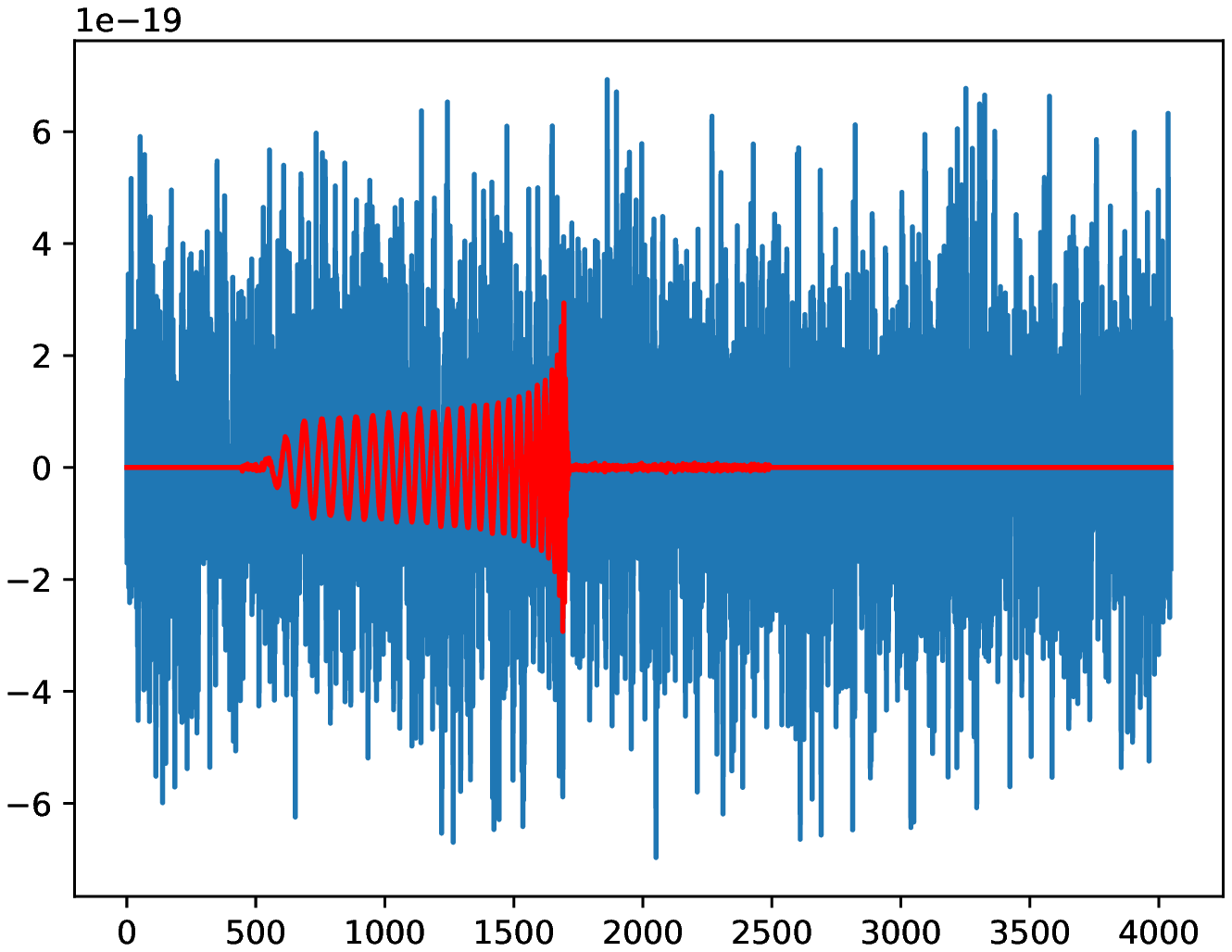]{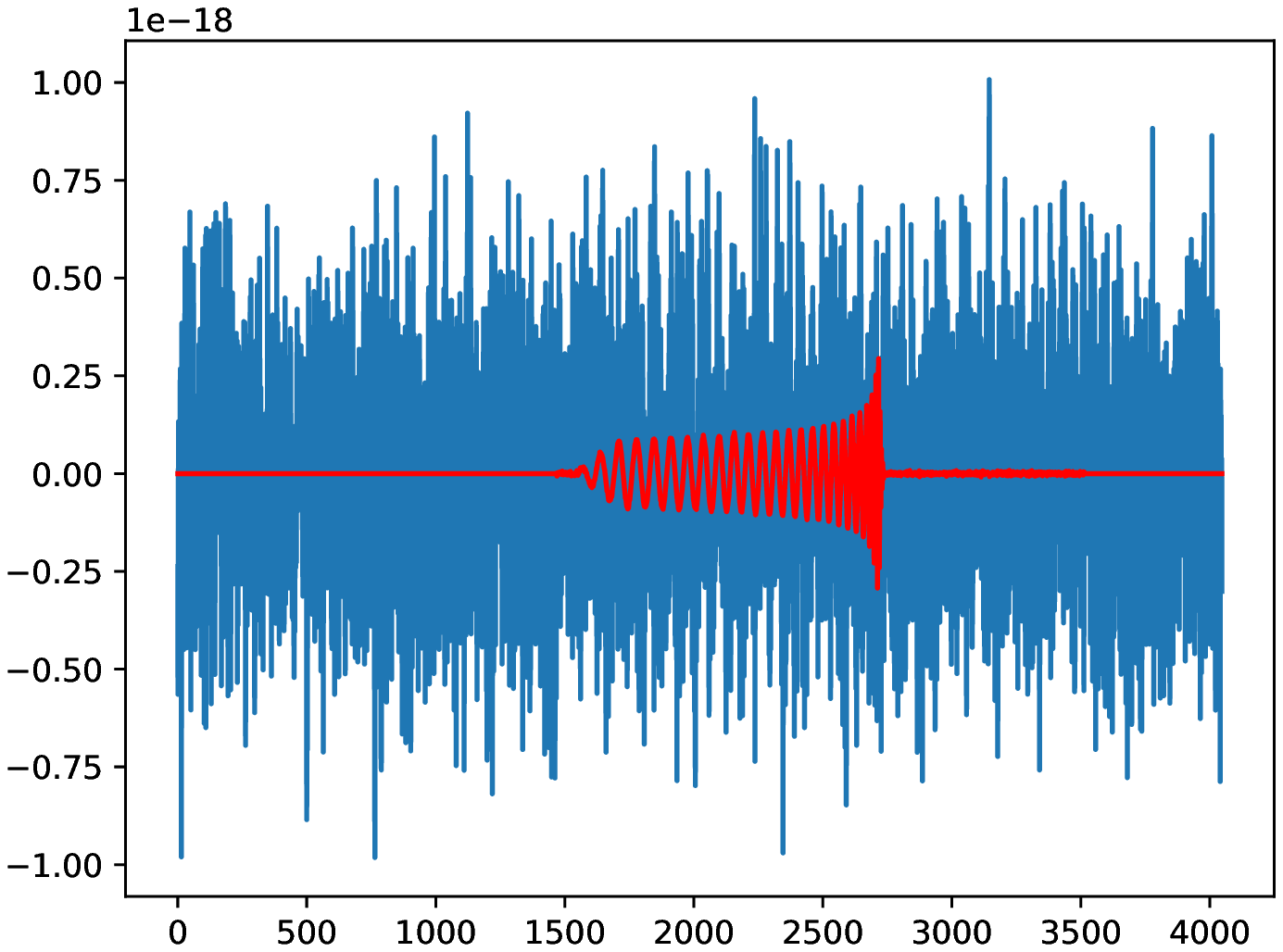}{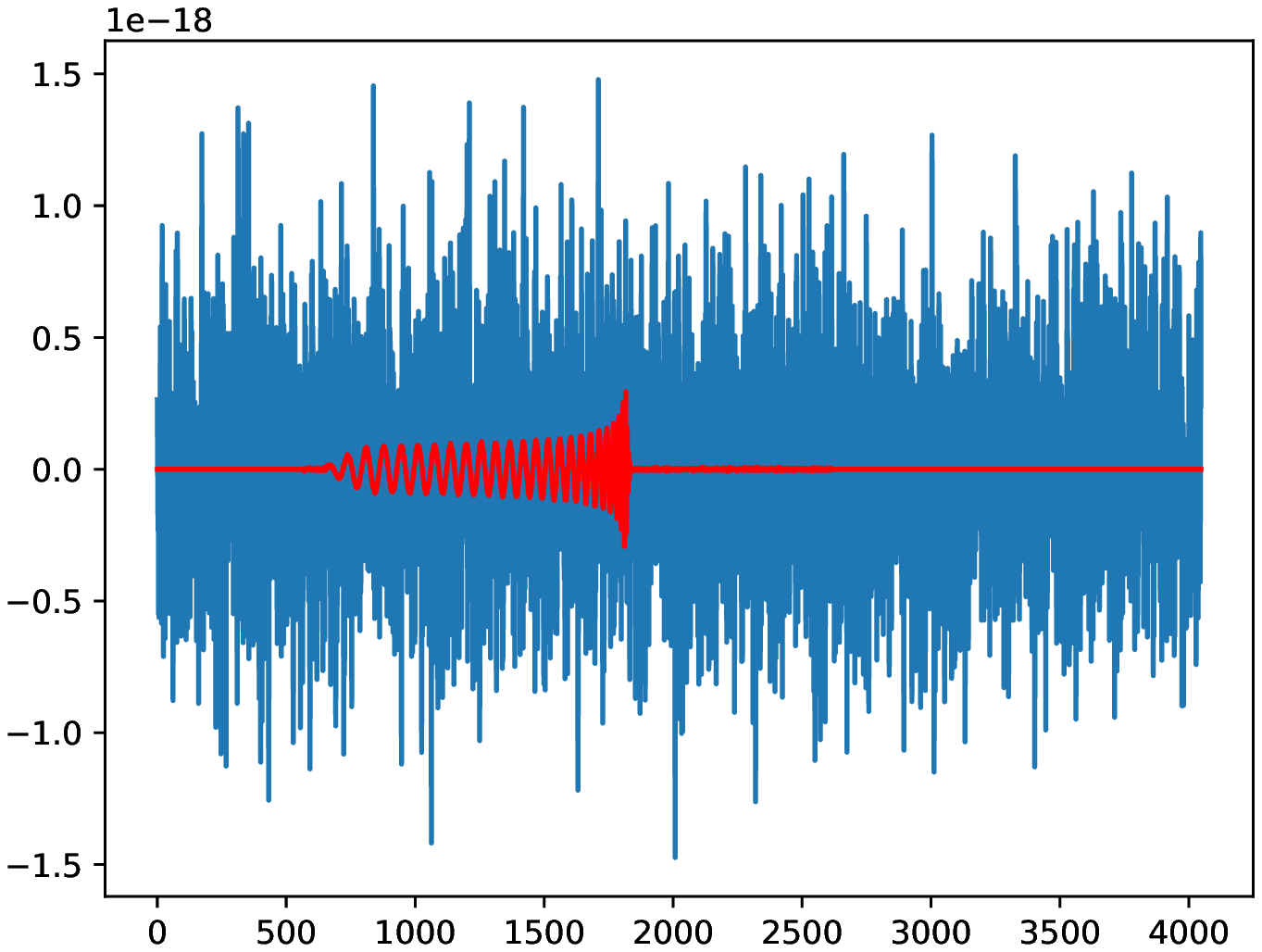}{0.32}{ Signals superimposed in red on noise with signal in blue. Random padding included. SNR: $12.519$,$9.782$,$6.874$ left to right, respectively.}{fig:sigs}

Figure \ref{fig:sigs} shows noisy signals (blue) that contain the GW  (red) with SNR = $12.519, 9.782, 6.874$, respectively. As shown in the figure, the GW is highly weak compared to the noise, so the detection is challenging. 

Note that the  noise is not Gaussian in general. It comes from various sources, terrestrial and astrophysical, giving color to the noise \cite{gwnoise,gwnoise2}.  Also note that SNR can be higher than the SNRs in the range used for our synthetic data \cite{ligo3,ligo33}. We focus on lower SNRs as we are interested in the limitations of the detection methods, though a SNR of below $10$ is generally considered too low to be verifiable. Most detected signals have a SNR within the range we chose to test at the detection sites \cite{ligo2,ligo4,ligo5,ligo1}.

\section{Sliding window embedding}
We use the persistent homology of sliding window embeddings described in \cite{sliding}  (see Section \ref{sec:hom} for persistent homology).  The main difference between the noise signal with and without the GW signal can be  characterized by the periodic embedding of signals. 

For a time-series $f_k, \hspace{2mm} k \in \{1, \cdots, N\}$ the sliding window embedding of size $m$ at the time index $j$ is:
\begin{eqnarray}
( f_j, f_{j+1}, ... f_{m+j-1} ),
\end{eqnarray}
where $N$ is the number of samples and there are  $N-m+1$ points in $\mathbb{R}^m$. 
A periodic signal has a sliding window embedding of a circle or an oval. A decaying periodic signal is a spiral. White noise is a ball. They are easily classified using homology groups, providing a method to classify different periodic-like behaviors \cite{sliding}.  

We chose $m=200$, which is not optimized but chosen heuristically.  Figure \ref{fig:embed} shows the sliding window of white noise with SNR $= 12.519, 9.782, 6.874$.

\threefigl[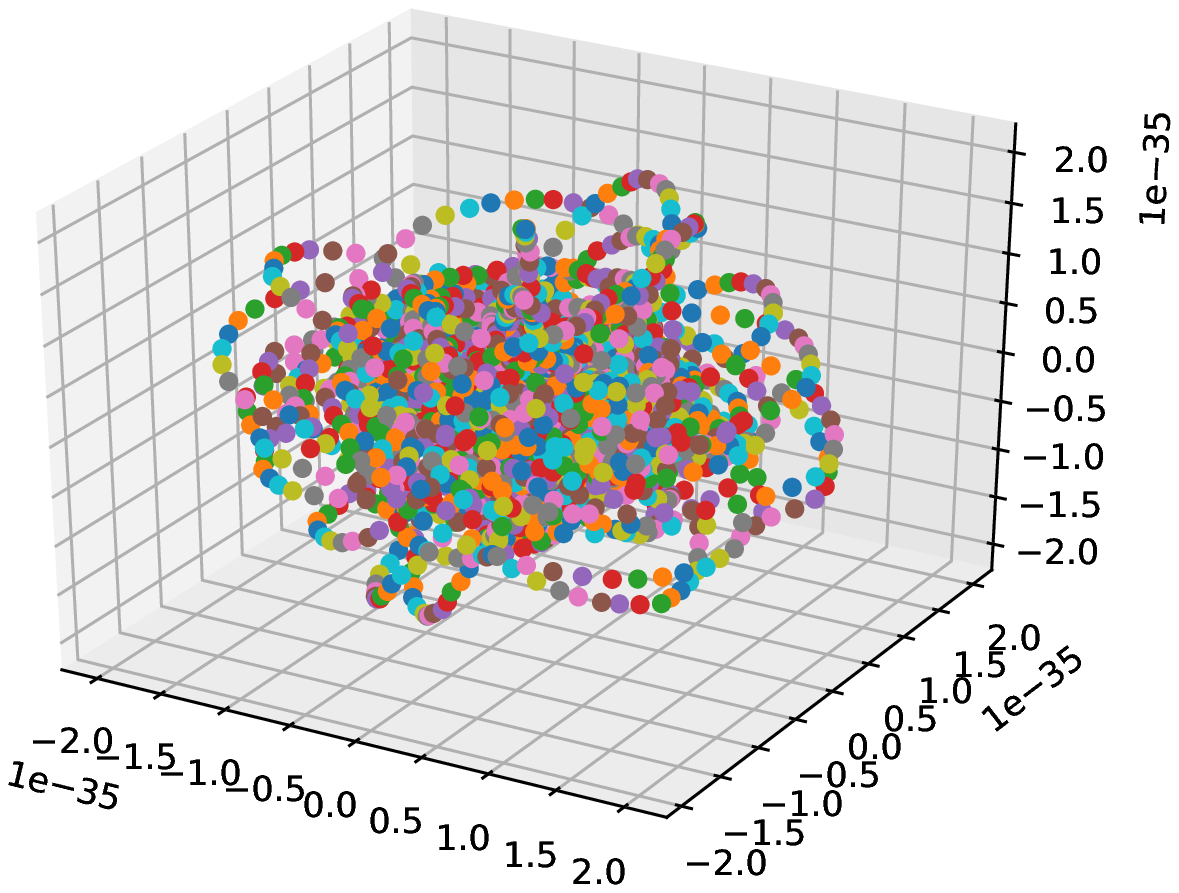]{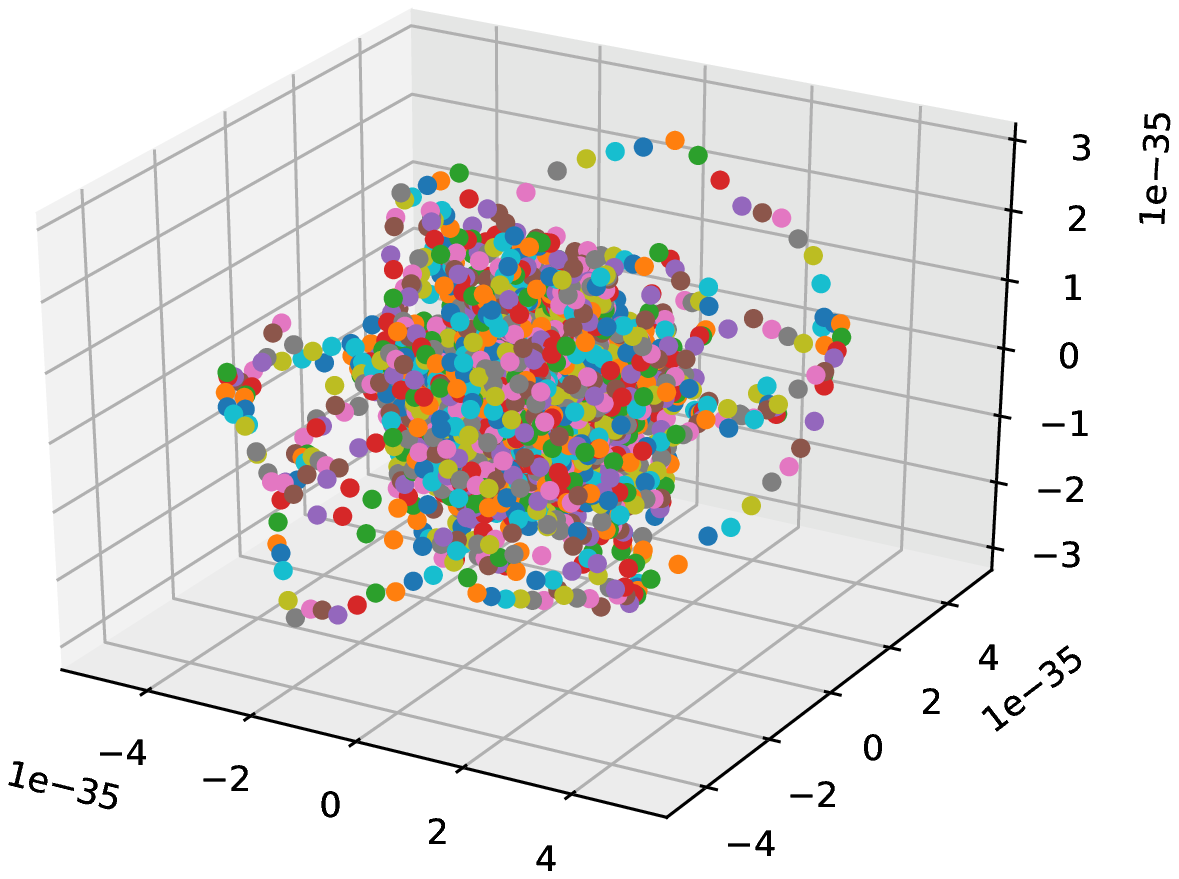}{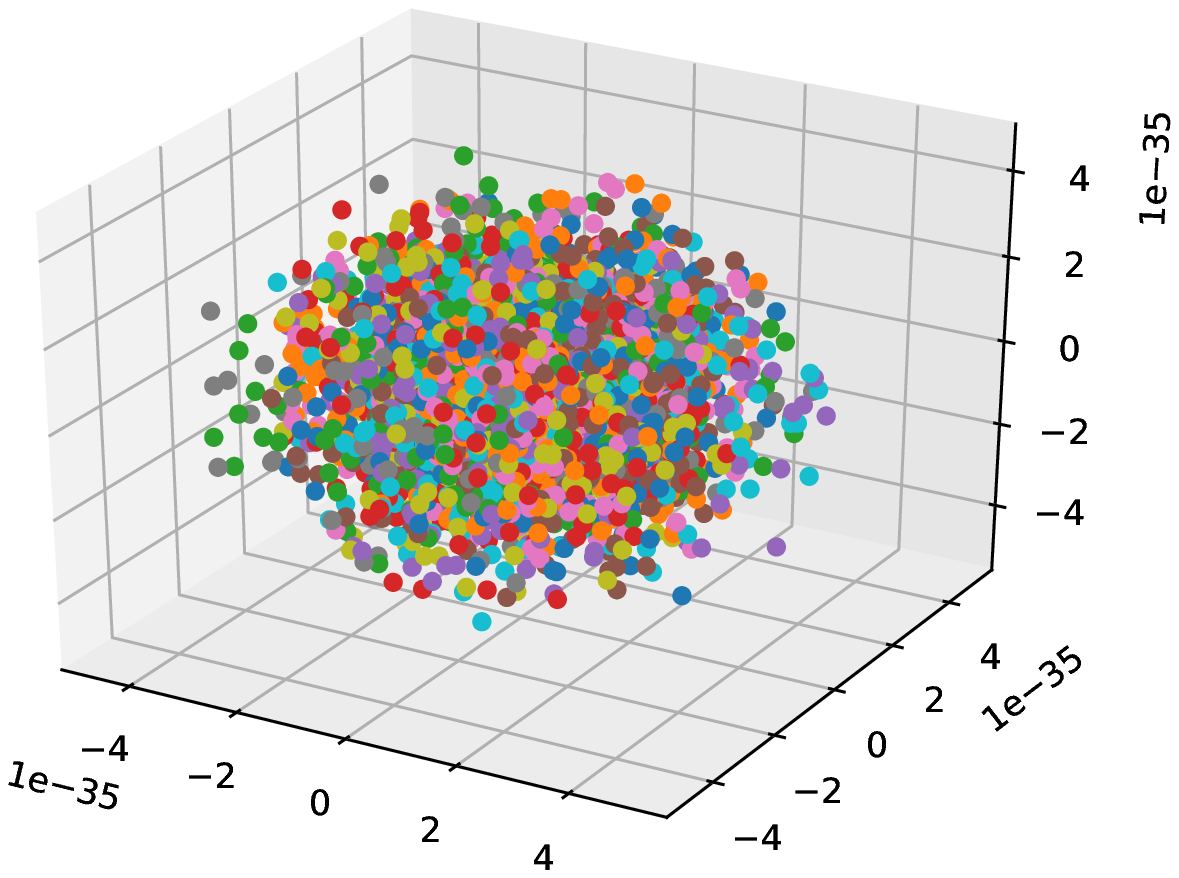}{0.32}{Sliding window with noise, SNR descending left to right $12.519$,$9.782$,$6.874$, respectively.}{fig:embed}

{\bf{Dimensional reduction}}: 
The sliding window creates a point cloud in a $m$-dimensional space. However, the features of interest are much lower dimensional. We calculate feature persistent homology of $H_1$, requiting $2$ dimensions. This, along with the expense of TDA on high dimensional space, motivates a dimensional reduction before proceeding. %does not require, just affords for, motivates, 
We project down to three dimensions for ease of visualization and to avoid loss of using only two. 

With the principal component analysis we project the data onto the first three singular vectors. Let $M$ be a matrix containing each window of length $m=200$ as a row. The point cloud is first centered at $0$:
\begin{eqnarray}
M_{:,j} = M_{:,j} - {\overline{M_{:,j}}},  \hspace{3mm} j \in \{0,1,\cdots, m-1\}, 
\end{eqnarray}
where $M_{:,j}$ denotes the usual matrix element notation and  ${\overline{M_{:,j}}}$ the mean value of $j^{th}$ column vector of $M$.  Then the singular value decomposition (SVD) is conducted and a projection operator is constructed to project the point cloud onto the first three scaled left singular vectors:
\begin{eqnarray}
U \Sigma V^* &=& M, \\
\tilde{M} &=& M(\Sigma_{1:3,1:3}V^*_{1:3,:})^*, 
%return dat*(np.diag(S[0:3])*V[0:3,:]).T
\end{eqnarray}
where $U\Sigma V^*$ is the SVD of $M$, ${\tilde{M}}$ the reduction of $M$ and the superscript $*$ the Hermitian conjugate. 

Figure \ref{fig:cloud} is the dimensionally reduced sliding window embedding of a GW signal (left) and white noise (right). Notice that the topology of the white noise is approximately a ball, while the GW chirp signature has a different topology. 

%Clearly the GW signals are topologically different from noisy signals. 

%Figure \ref{fig:cloud} visualizes the sliding window of the GW signals after the dimensional reduction (left) and that of the noise (right). Notice that the white noise is more like a random ball. Clearly the GW signals are topologically different from noisy signals. 
%%%%% HERE HERE HERE 
\twofigl[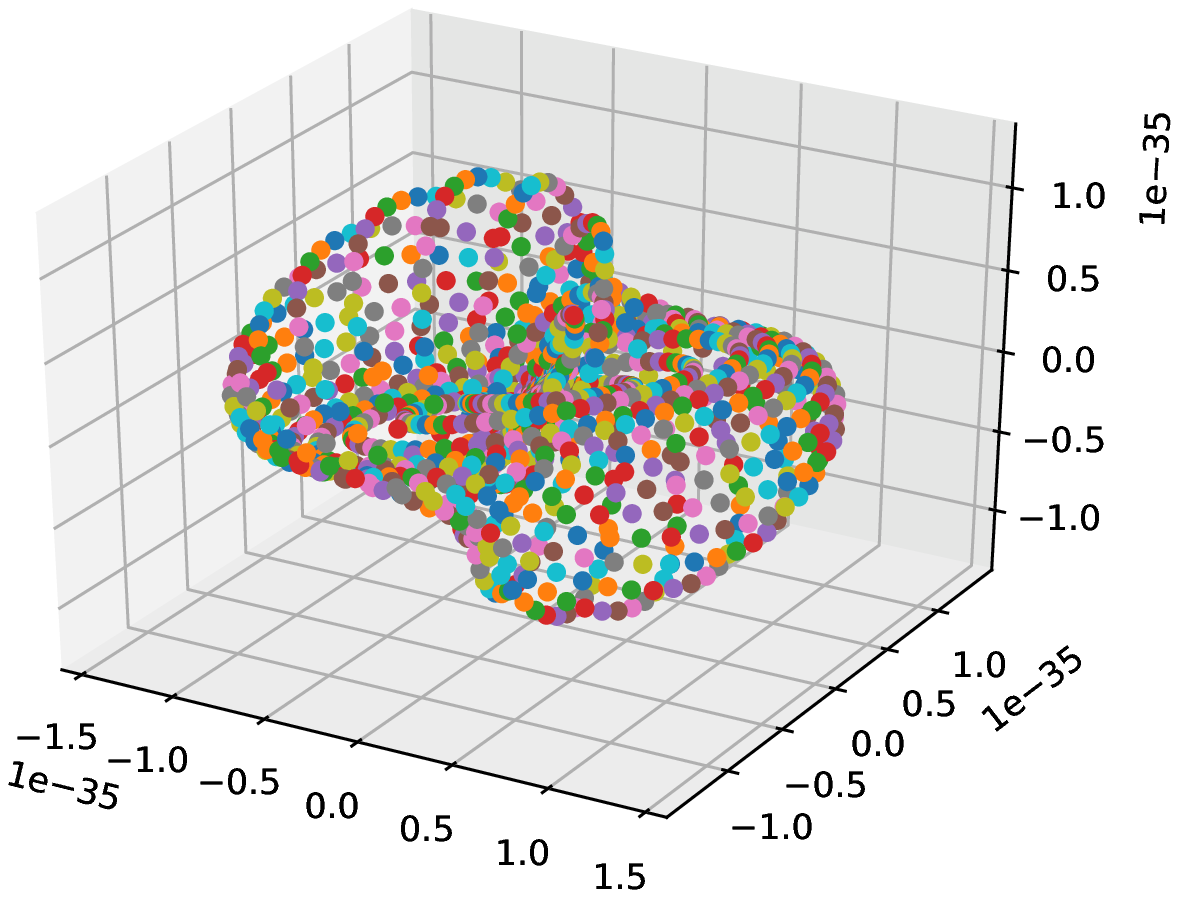]{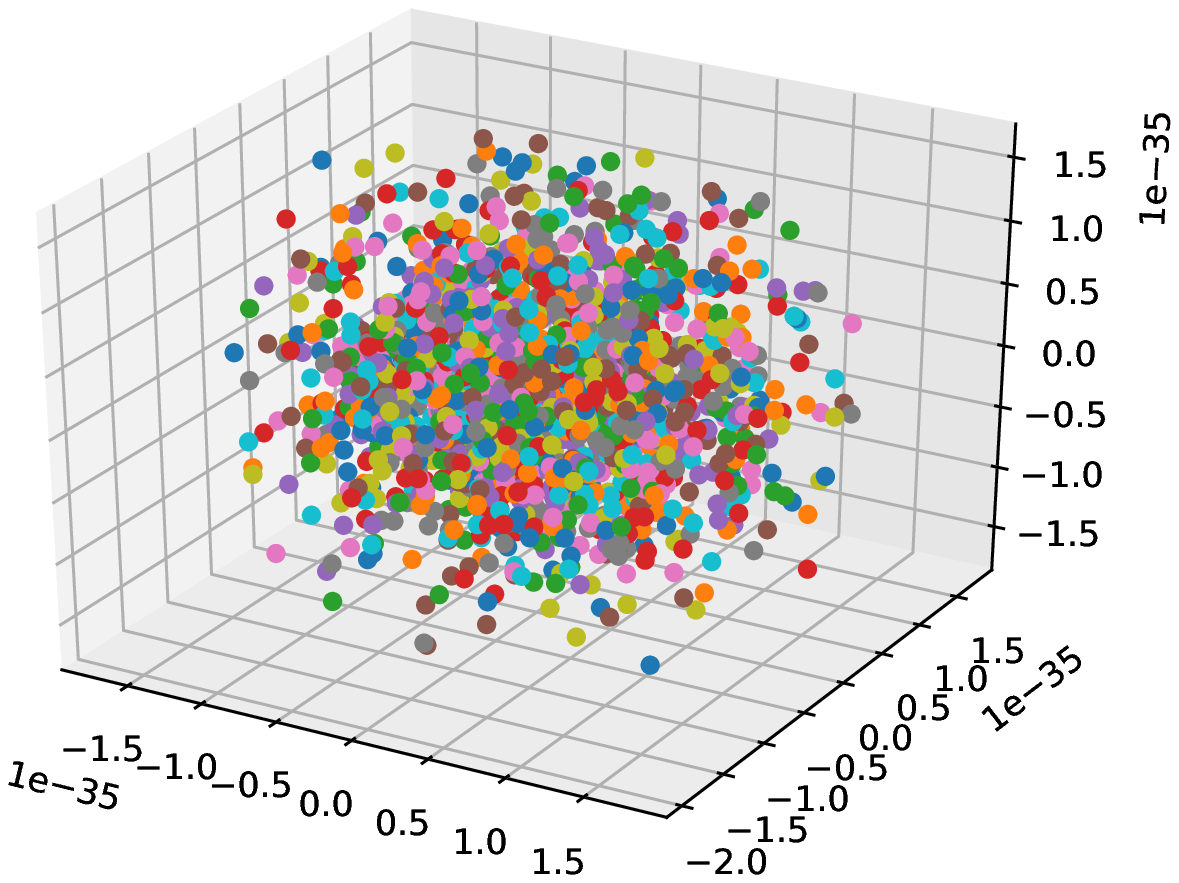}{0.45}{Left: Sliding window of a GW signal with $m=200$ after the dimensional reduction to $3$. Right: Sliding window embedding of white-noise.}{fig:cloud}

%%% INSERTED TDA SECTION HERE

\section{TDA: Persistent Homology}\label{sec:hom}
Let $X$ be the topological space of interest, the embedding space of GW signals in our case. Let $\sigma_n$ be the $n$-simplex, the convex hull composed of $n+1$ vertices. 
Let $\mathbf{R}$ be a ring and $C_n(X,\mathbf{R})$ be the free $\mathbf{R}$-{module} generated by all possible continuous images of $n$-{simplices} $\sigma_n$. Let $\delta_n$: $C_n$ $\rightarrow$ $C_{n-1}$ be the boundary map as
$$
    \delta_n\sigma_n = \sum_{k=0}^n (-1)^k \left[p_0, \cdots, p_{k-1},p_{k+1},\cdots,p_n \right]
$$
where $\left\{ p_i \right\}$ are the vertices. The factor $(-1)^k$ is put to preserve the orientation. The $n$th homology group of $X$ with coefficients in $\mathbf{R}$, $H_n(X,\mathbf{R})$, is the quotient group of the kernel and image groups:
$$
H_n(X,R) = ker(\delta_n)/im(\delta_{n+1}).
$$
%where $ker$ and $im$ denote the kernel and image groups. 
%For a given topological space, the $n\textit{th}$ homology group generally refers to the number holes in the $n$th dimension. 
The number of generators of $H_n(X,\mathbf{R})$ is called the Betti number, $\beta_n$, roughly the number of geometric holes in $n$th dimensional space of $X$. $\beta_0$ denotes the number of connected components. 
For example, for $S^2$, $H_0 = 1$, $H_1 = 0$ and $H_2 = 1$ and for $T^2$, $H_0 = 1$, $H_1 = 2$ and $H_2 = 1$. 

Given the point cloud, we construct a simplicial complex created by gluing a finite number of simplices together.
%A simplex is created by first taking some 0-simplices and attaching edges to pairs of points. Three adjacent edges may be filled in with a triangle, and four adjacent triangles may be filled in by a tetrahedron, etc. Note that edges and triangles may not intersect except at their vertices/edges, etc. 
Homology on such a simplicial complex is known as persistent homology.  We use the Vietoris-Rips complex \cite{topologyanddata}, which  is built by taking all  points as zero simplices. For a fixed value of  $t$, known as the filtration parameter,   we add edges between two points if their distance is less than $t$. We add a triangle between three points if each pair of points has distance less than $t$ and so on for higher dimensional simplices. We repeat this process with various $t$ values.

The barcode is the graph of $\beta_n$  against the parameter $t$.  It displays not only $\beta_n$ at each $t$, but also graphs how long each generator remains non-trivial. Its interval gives a concept referred to as {\it persistence}. The starting point of each persistence is called ``birth" and the ending  ``death". Its vertical representation, with the brith as $x$-axis and the death $y$-axis, is called the persistence diagram. 

Let $l_0$ be the number of the persistences in $H_0$ barcode and $l_1$ in $H_1$. Once both barcodes are obtained, we sort the persistences by descending magnitude. Let $\pi^0_i$ be the ordered persistence in $H_0$, $i = 1, 2, \cdots, l_0$, similarly $\pi^1_i$ the ordered persistence in $H_1$, $i = 1, 2, \cdots, l_1$. Let  $\Pi^0$ and $\Pi^1$ be 
\begin{eqnarray}
\Pi^0 &=& (\pi^0_1, \pi^0_2, \cdots, \pi^0_{l_0}), \nonumber \\
\Pi^1 &=& (\pi^1_1, \pi^1_2, \cdots, \pi^1_{l_1}).  \nonumber 
\end{eqnarray}
We call these persistence vectors, and they are how we choose to encode topological features as a vector that can be used as input to a CNN. 
%These vectors are topological features to be combined with CNN in the following. 

%We use the persistent homology for dimensions $0$ and $1$. The extension to $H_2$ is worthy of  investigation, which can contain more topological features.  $H_0$ was included as it was available and the nature of a CNN allows for it to be used or disregarded as fit. The Vietoris-Rips complex is computationally feasible as we use only $H_0$ and $H_1$ but may become resource intense for higher dimensional homology. 

We use the persistent homology for dimensions $0$ and $1$  because the existing results regarding TDA of sliding window embeddings focus on $H_1$ \cite{sliding}. $H_0$ was included as the performance impact of doing so was negligible and the nature of a CNN allows for it to be used or disregarded as fit. The Vietoris-Rips complex is computationally feasible as we use only $H_0$ and $H_1$ but may become extremely resource intense for higher order homology groups.

\section{Preprocessing for CNN}

The persistence vectors $\Pi^0,\Pi^1$ (of length $l_0,l_1$ respectively) are adjusted to a fixed length $N_p$ by either truncation or zero-padding as needed. These are then concatenated into one vector $\Pi$, of length $2N_p$:
\begin{eqnarray}
     \Pi = (\Pi^0, \Pi^1). \label{eq:tda}
\end{eqnarray}
The raw signal is then concatenated with $\Pi$:
\begin{eqnarray}
	\bm{x} = (signal, \Pi).  \label{eq:cat}
\end{eqnarray}
The resulting vector $\bm{x}$ is of fixed size and ready for input into the CNN. In the actual procedure, we first normalize $\Pi$ and $signal$ separately so that they have the same maximum. A similar approach has been applied to VLBI signal analysis \cite{Dongjin}.

\section{Hyper-parameters \& Procedure}
The hyper-parameters used here are meant to replicate the work in \cite{huerta}. They are suboptimal but we use them for comparison purpose. 
\begin{table}[b]
\begin{ruledtabular}
\caption{\label{tab:table4} The hyperparameters are used to replicate the work in \cite{huerta}. }
\begin{tabular}{c| c c}
	Number & Type & Parameters \\
	\hline
	1 & Input &  \\
	2 & Convolution & 64, strides = 1, kernel size = 16 \\
	3 & Max Pooling & strides = 4, pool size = 4 \\
	4 & Dense & 64, ReLU \\
	5 & Convolution & 128, strides = 1, kernel size = 16 \\
	6 & Max Pooling & strides = 4, pool size = 4 \\
	7 & Dense & 128, ReLU \\
	8 & Convolution & 256, strides = 1, kernel size = 16 \\
	9 & Max Pooling & strides = 4, pool size = 4 \\
	10 & Dense & 256, ReLU \\
        11 & Convolution & 512, strides = 1, kernel size = 32 \\
        12 & Max Pooling & strides = 4, pool size = 4 \\
        13 & Dense & 512, ReLU \\
	14 & Flatten & \\
	15 & Dense & 128, Linear \\
	16 & Dense & 128, ReLU \\
	17 & Dense & 64, Linear \\
	18 & Dense & 64, ReLU \\
	19 & Dense & 2, Linear \\
%\hline

\end{tabular}
\end{ruledtabular}
\end{table}
We used mean-squared error for the loss function and Adam optimizer \cite{adam} for optimization. Five epochs were used for training. The size of the synthetic data  sets is  $30,000$. This was divided into $20,000$ elements for training and $10,000$ for testing. Initialization function was \verb|Orthogonal()| in Keras. Random seeds were fixed everywhere necessary to force deterministic initialization and optimization for reproducibility.  
%\section{General process}
The procedure is as follows: 
%\begin{enumerate}
%\item Generate sliding window embedding of the raw signal
%\item Perform the dimensional reduction on the sliding window embedding, yielding a 3-dimensional point cloud
%\item Compute persistent homology in $H_0$ and $H_1$ of the sliding window embedding
%\item Process the persistence intervals into fixed length real vectors% as the lengths of all finite intervals ordered descending, padding with $0$'s and/or truncating to a fixed length.  
%\item Normalize the persistence vectors and raw data so they have the same maximum
%\item Concatenate the normalized raw signal and persistence vectors
%\item Input into CNN for binary classification. 
%\end{enumerate}

\vskip .1in 

\noindent
{\it{1. Generate sliding window embedding (SWE) of the raw signal}}

\noindent
{\it{2.  Perform the dimensional reduction on the SWE, yielding a $3$-dimensional point cloud}}

\noindent
{\it{3. Compute persistent homology of $H_0$ and $H_1$ of the SWE}}

\noindent
{\it{4. Construct  $\Pi$ with a fixed $N_p$}}% as the lengths of all finite intervals ordered descending, padding with $0$'s and/or truncating to a fixed length.  

\noindent
{\it{5.  Normalize $\Pi$ and raw data}}

\noindent
{\it{6. Concatenate $\Pi$ and the raw signal}}

\noindent
{\it{7. Input into CNN for binary classification}} 

\section{Results}

{\bf{Performance metrics}}:
{\it Sensitivity} is the ratio of true positives to all positives and {\it specificity} is the ratio of true negatives to all negatives. They are also called the true positive rate (TPR) and true negative rate (TNR), respectively. A perfect classifier has both equal to $1$. The case of $0.5$ is equivalent to using a coin toss as a binary classifier. The case where TPR  $=1$ and TNR  $=0$ corresponds to a case where the classifier always guesses positive regardless of input, vice versa if the classifier always guesses negative.

The receiver operating characteristic (ROC) curves are another metric for the evaluation. The closer the area under the curve (AUC) to $1$, the better the classifier. A perfect binary classifier is a step function that reaches $1$ at $x=0$. A classifier of  $y=x$ shows no classification ability equivalent to a coin fair flip.  

{\bf{Software}}:
We used GUDHI \cite{gudhi,gudhi2} for TDA and Keras for CNN as an interface to Tensorflow. 
For the ROC curves \verb|sklearn.metrics.roc_curve()| was used. For the sensitivity and specificity vs SNR curves, our own routine was used with a fixed threshold value of $0.5$.   \verb|python-gwtools| was used to calculate the optimal match-filtered SNR \cite{gwtools}.

Figures \ref{fig:one1500}, \ref{fig:one}, \ref{fig:three} and \ref{fig:two} show two performance metrics for three different training and test sets. {{The blue solid line represents the CNN with raw data, the red the CNN with TDA features only  (Eq. \eqref{eq:tda})  and the green the CNN with raw data and TDA features concatenated (Eq. \eqref{eq:cat}), labeled as  \verb|raw, tda, both|.}} Each contains $20,000$ training elements, 50\% of which have a GW signal. The test set has $10,000$ elements synthesized in the same manner as the training set. The first set uses $11$ different signals of different mass ratios, as well as $10$ different SNRs. The other two sets use only one signal with $30$ uniform samplings and $100$ SNRs respectively, for $R \in [0.075,0.65]$.
%, see table \ref{tab:snr}.
The left % for figures ~\ref{fig:one1500},~\ref{fig:one},~\ref{fig:two},~\ref{fig:three} 
shows the sensitivity and specificity for each method versus SNR.

\twofigl[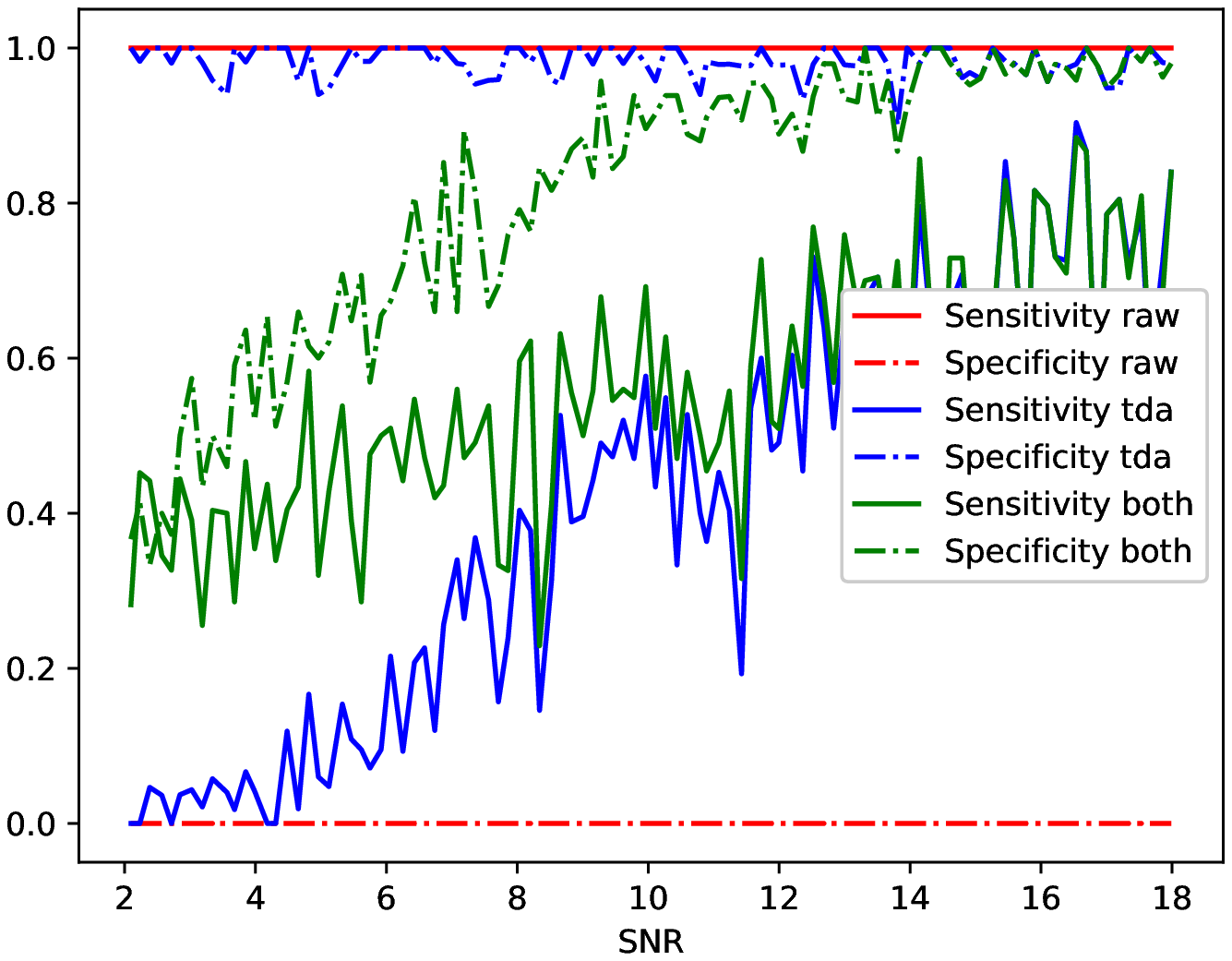]{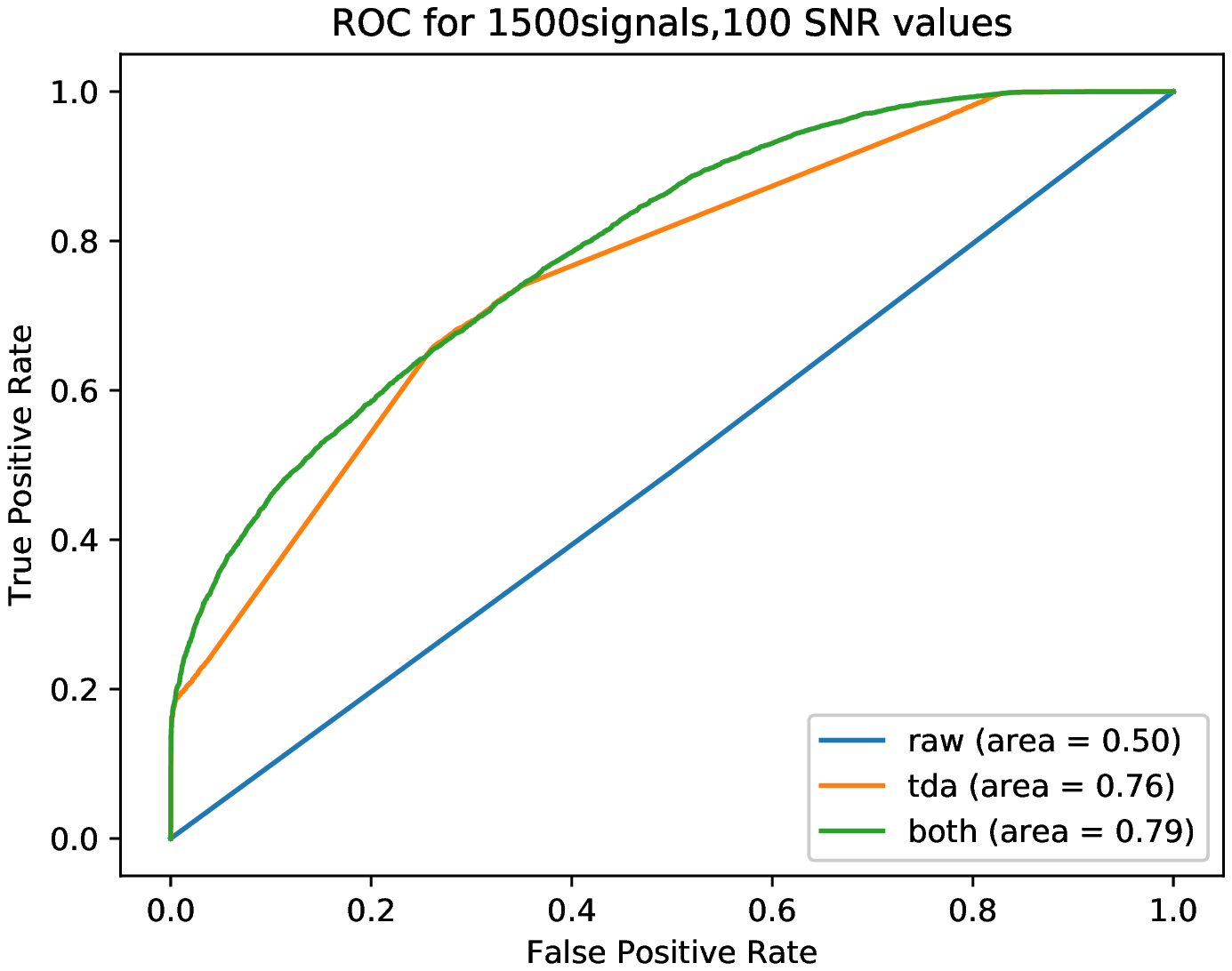}{0.49}{Hardest case, $1500$ different GW signals of mass ratio $1$ to $9.97$ sampled at $100$ different SNRs. Raw signal has no detection capability}{fig:one1500}

\twofigl[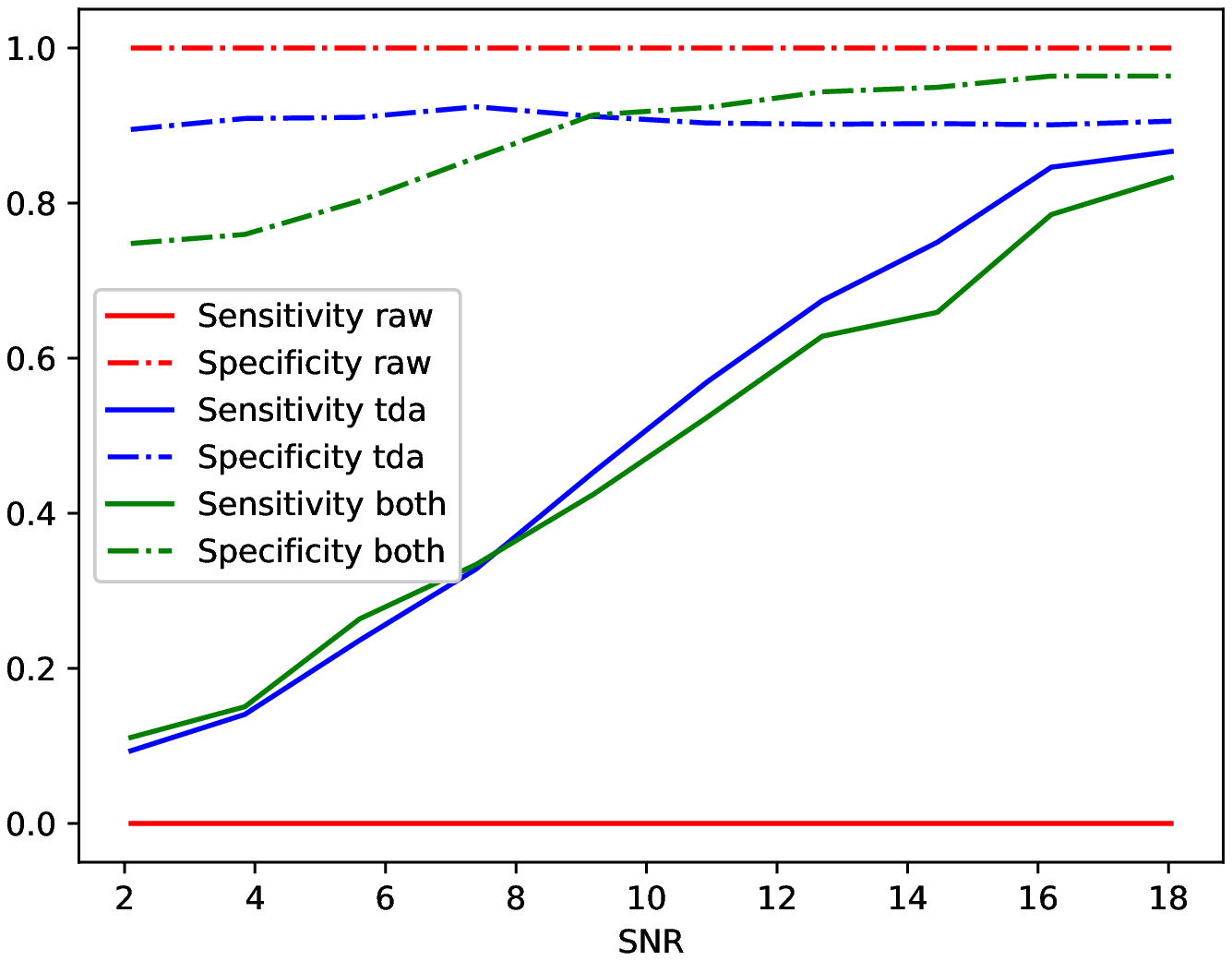]{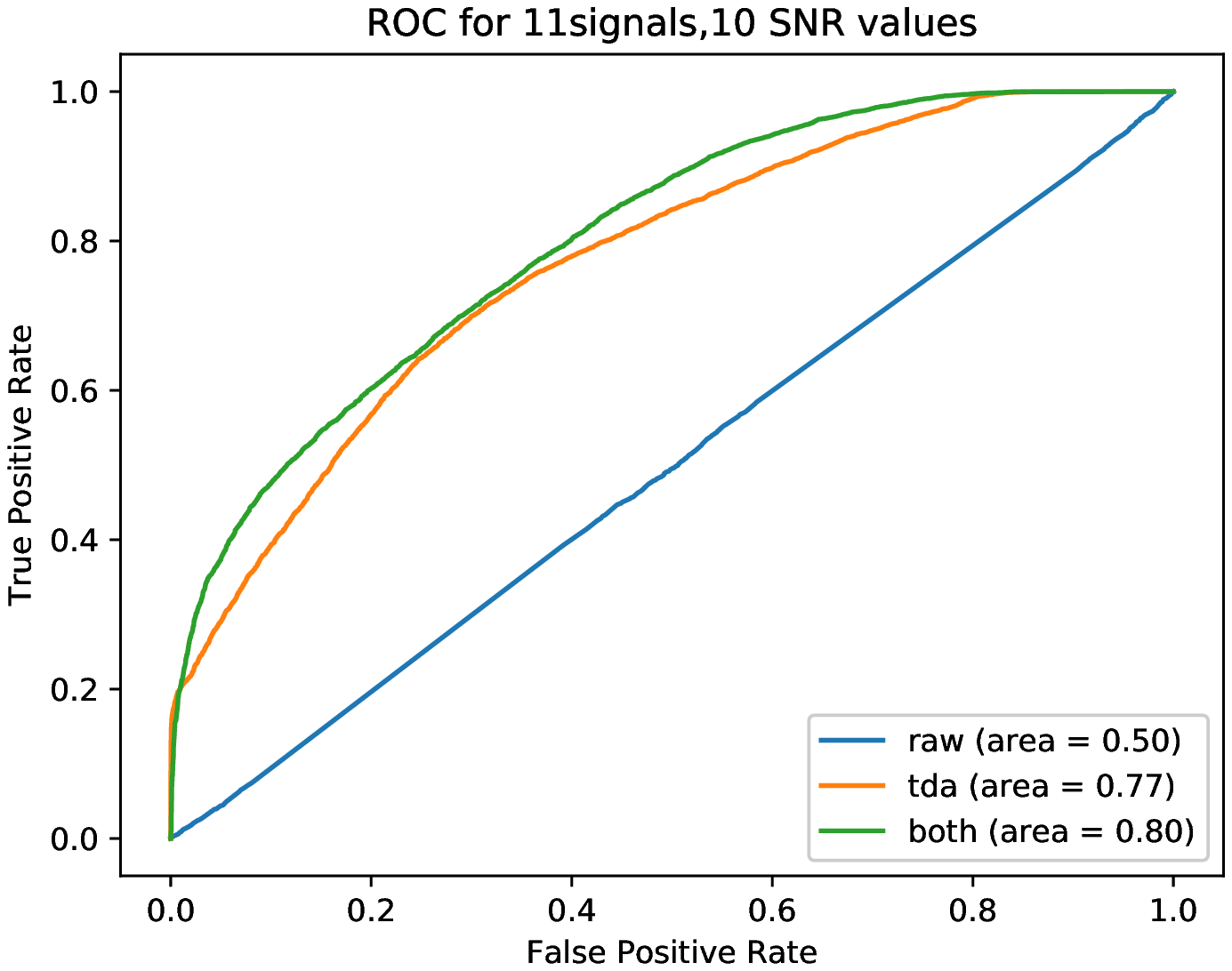}{0.49}{Hard case, $11$ different GW signals of mass ratio $1$ to $9.97$ sampled at $10$ different SNRs. Raw signal has no detection capability}{fig:one}

%\begin{figure}[h]
%\includegraphics[width=0.45\textwidth]{figs/sensspecRN30sigsamp1.eps}
%\caption{Random 30 different SNR samples, 1 signal $20000$ training set 50\% signal prescence. $10000$ test set}
%\end{figure}

\twofigl[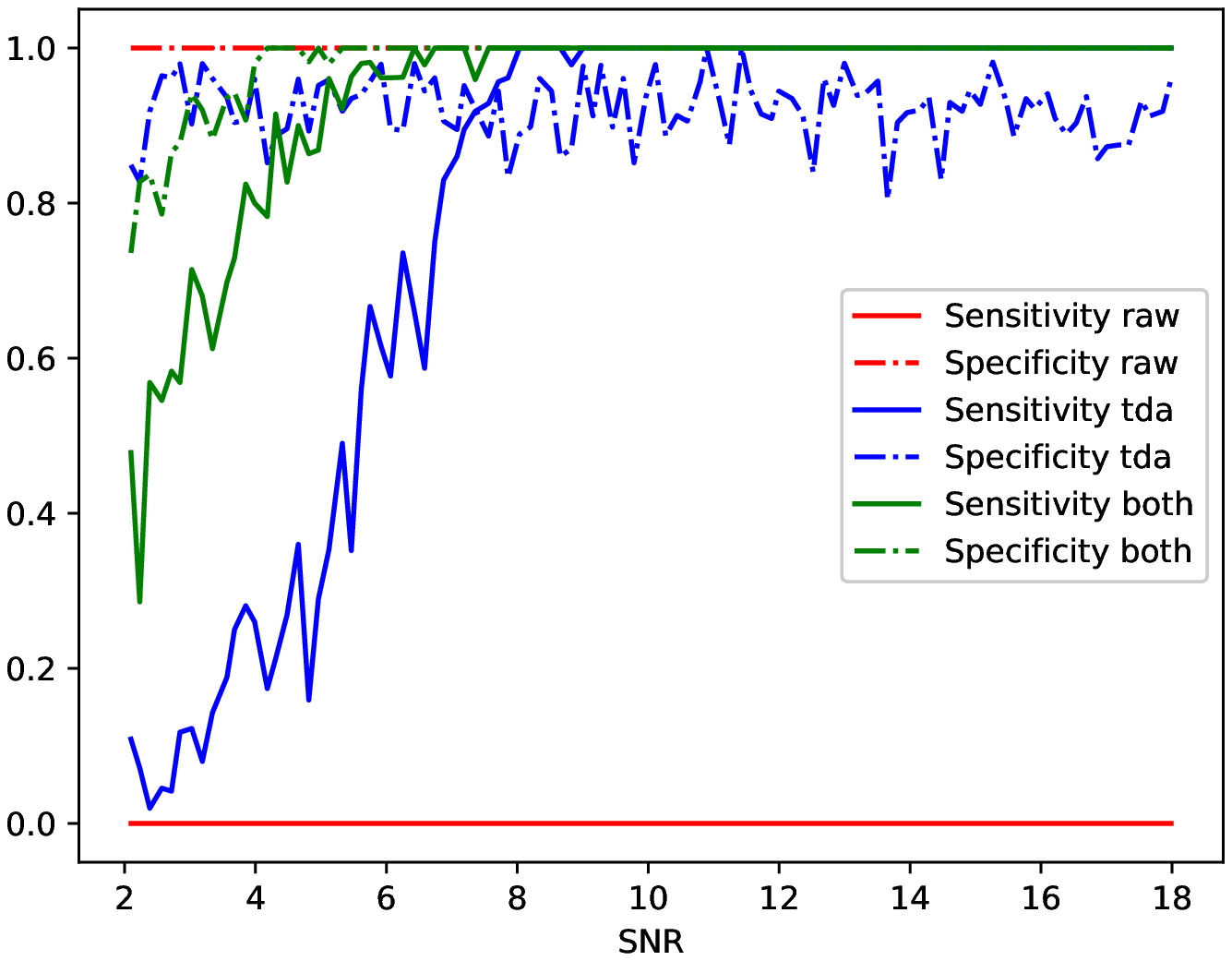]{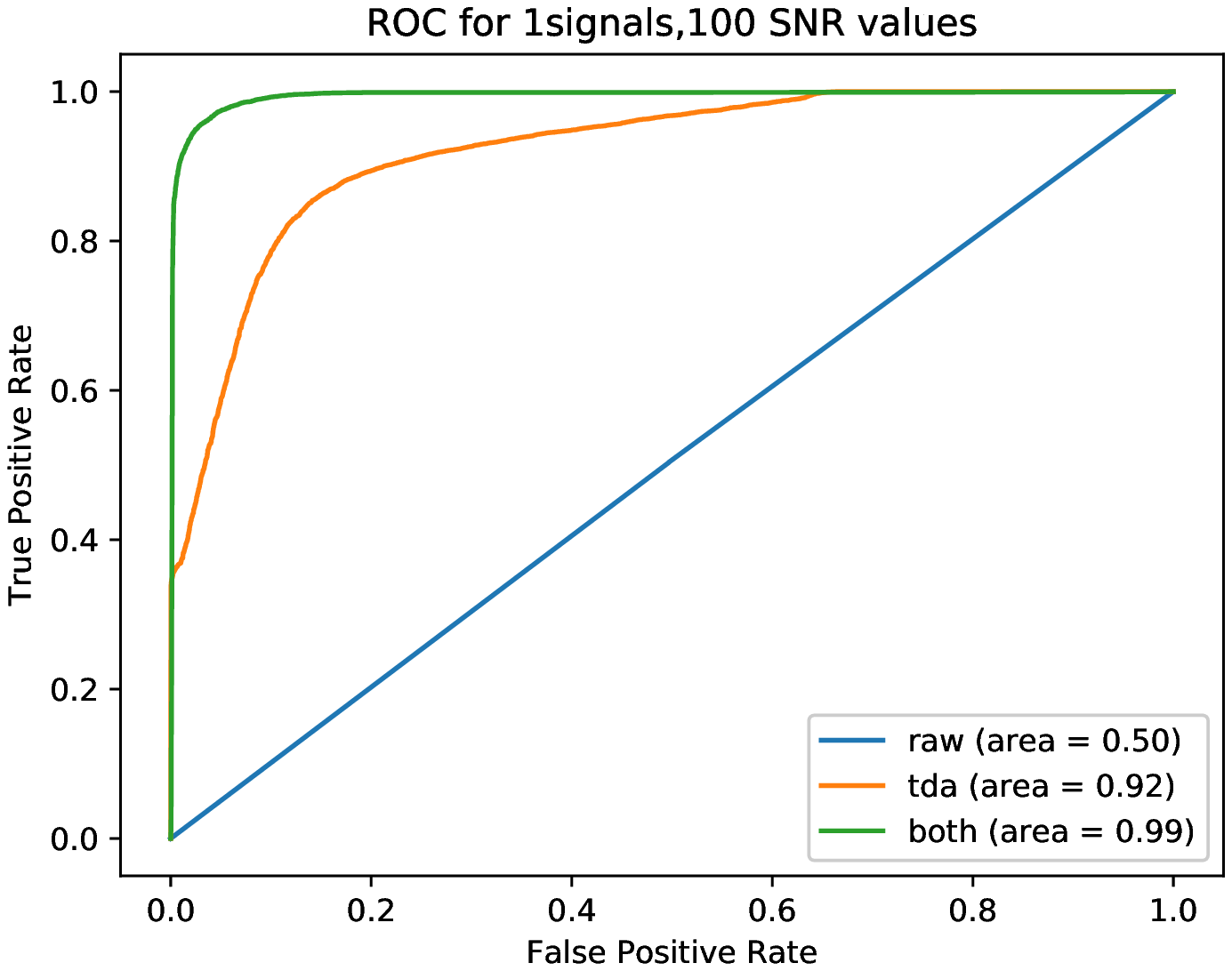}{0.49}{$100$ different SNRs. Raw signal alone has no detection capability.}{fig:three}

\twofigl[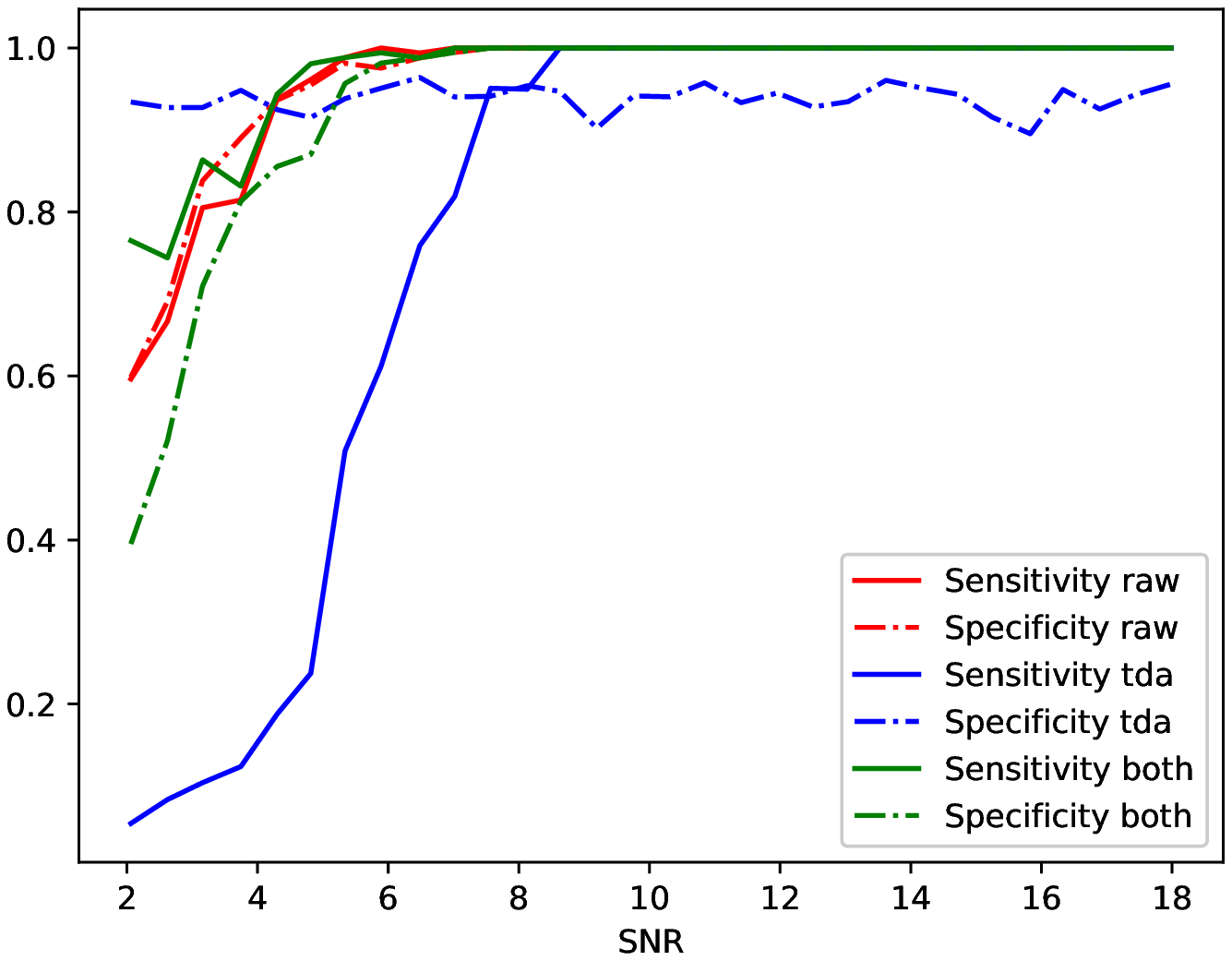]{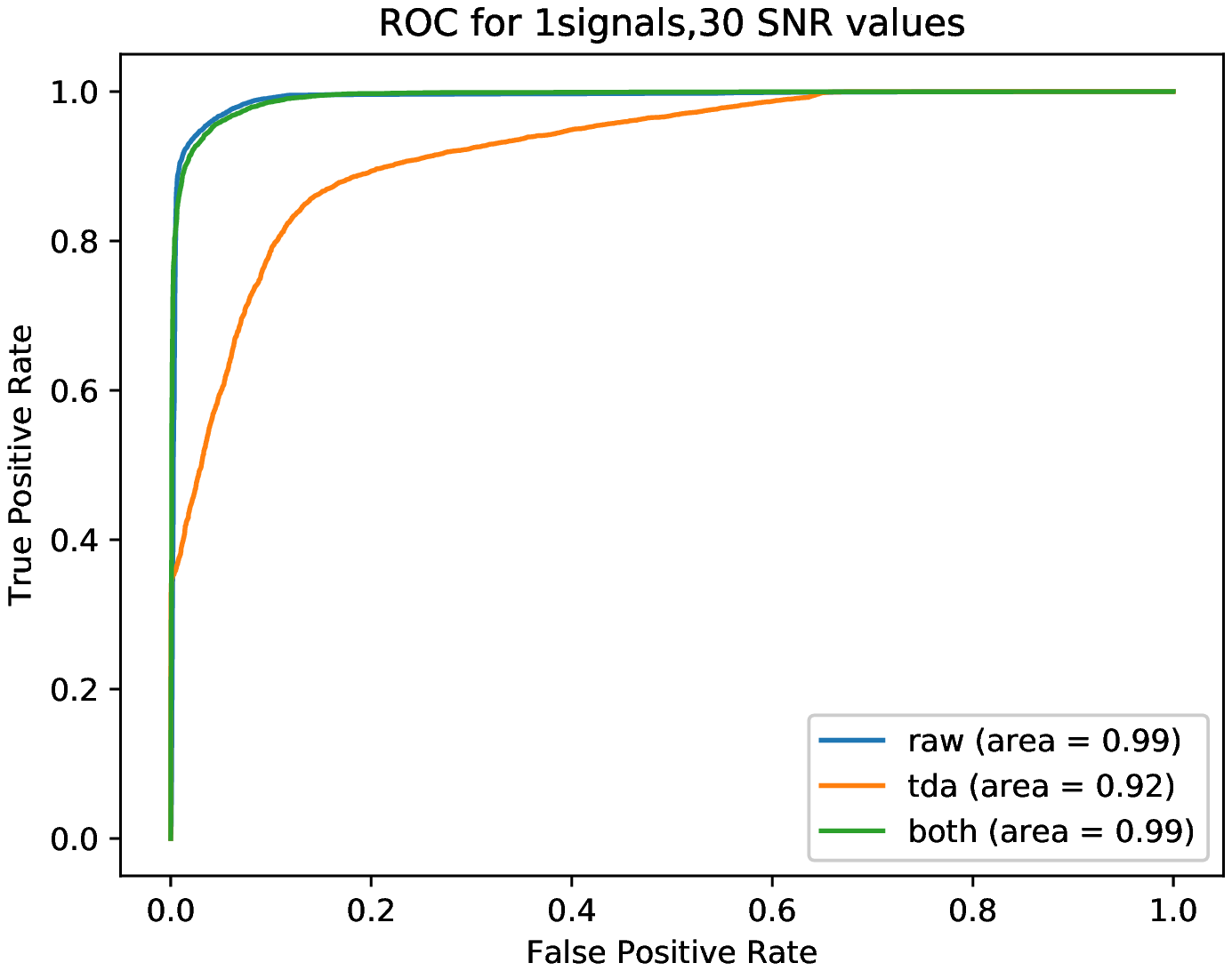}{0.49}{Easiest case, only $30$ different SNRs, $1$ signal. }{fig:two}

As detection is more difficult when there are a wider range of signals and SNR values (Fig. \ref{fig:one1500} and Fig. \ref{fig:one}), these tests are descending in difficulty, Fig. \ref{fig:two} being the easiest case and Fig. \ref{fig:one1500} being the hardest. It is clear that using the CNN with the raw signal alone provides no detection ability except for the easiest case (Fig. \ref{fig:two}), while the TDA features alone provide some detection ability in all cases, and the combined features are better in every case. This shows the power of the topological features to greatly improve performance, and the synergy of combining them with the raw signal (which increases maximum accuracy).

Figures \ref{fig:roc25}, \ref{fig:roc35},  \ref{fig:roc45} have a constant SNR and use $1500$ signals with mass ratios between $1.0078$ and $9.9759$. The training set sizes are $11000,4000,1500$ and test set sizes $4000,2000,1500$,  respectively. They show the effect of noise level and training set size on the efficacy of the method. They show that the TDA features are responsible for increased accuracy at lower SNR with less training. Note that when training the CNN with TDA features alone, performance does not change much as the training set size increases. It is worth noting that when using TDA features alone, performance maximizes after a relatively small training set. This lends to the asseration that the TDA features reduce the training requirements of the scheme. 

%This shows that the TDA features , as they reach maximum performance after a small amount of training.  

\threefigl[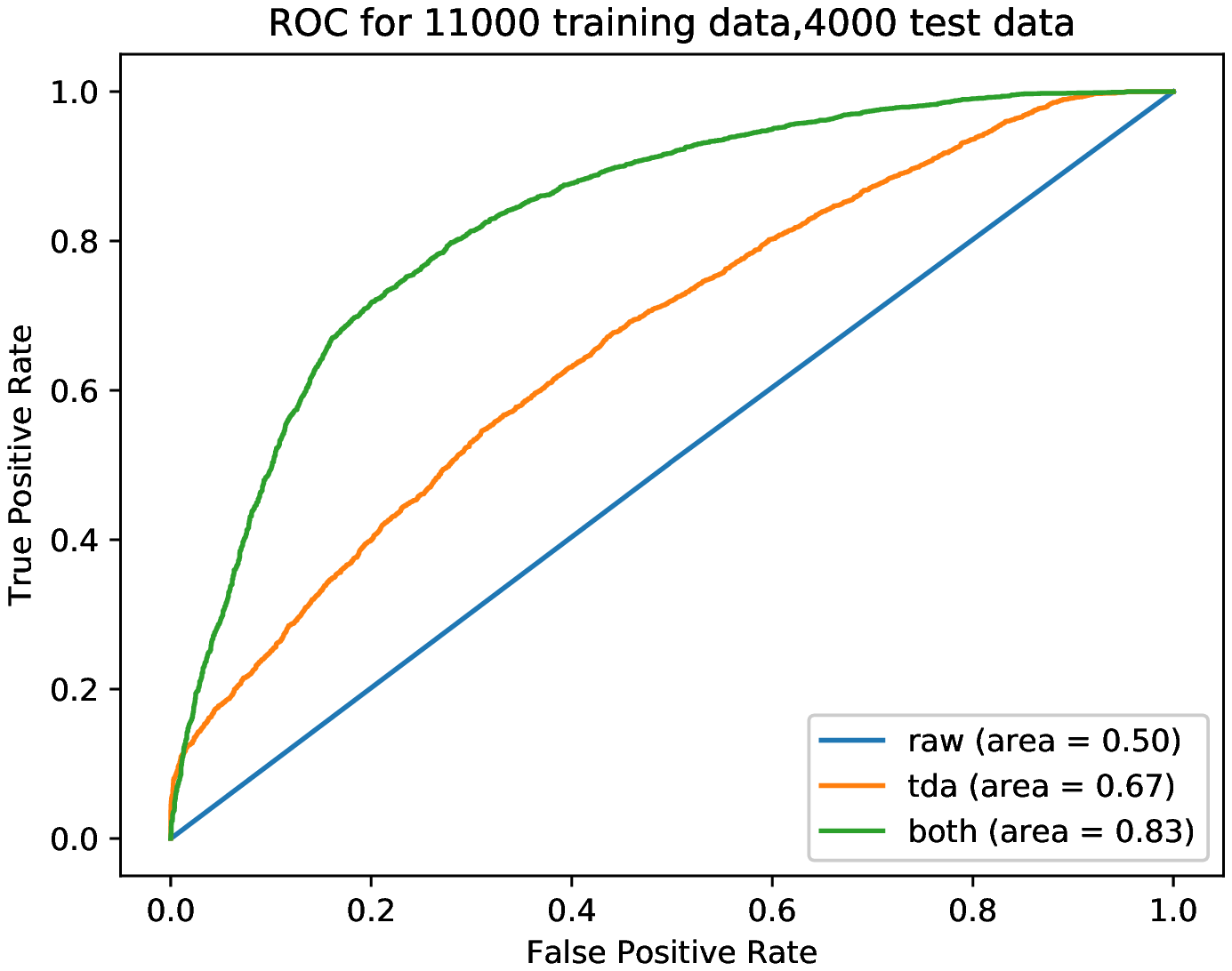]{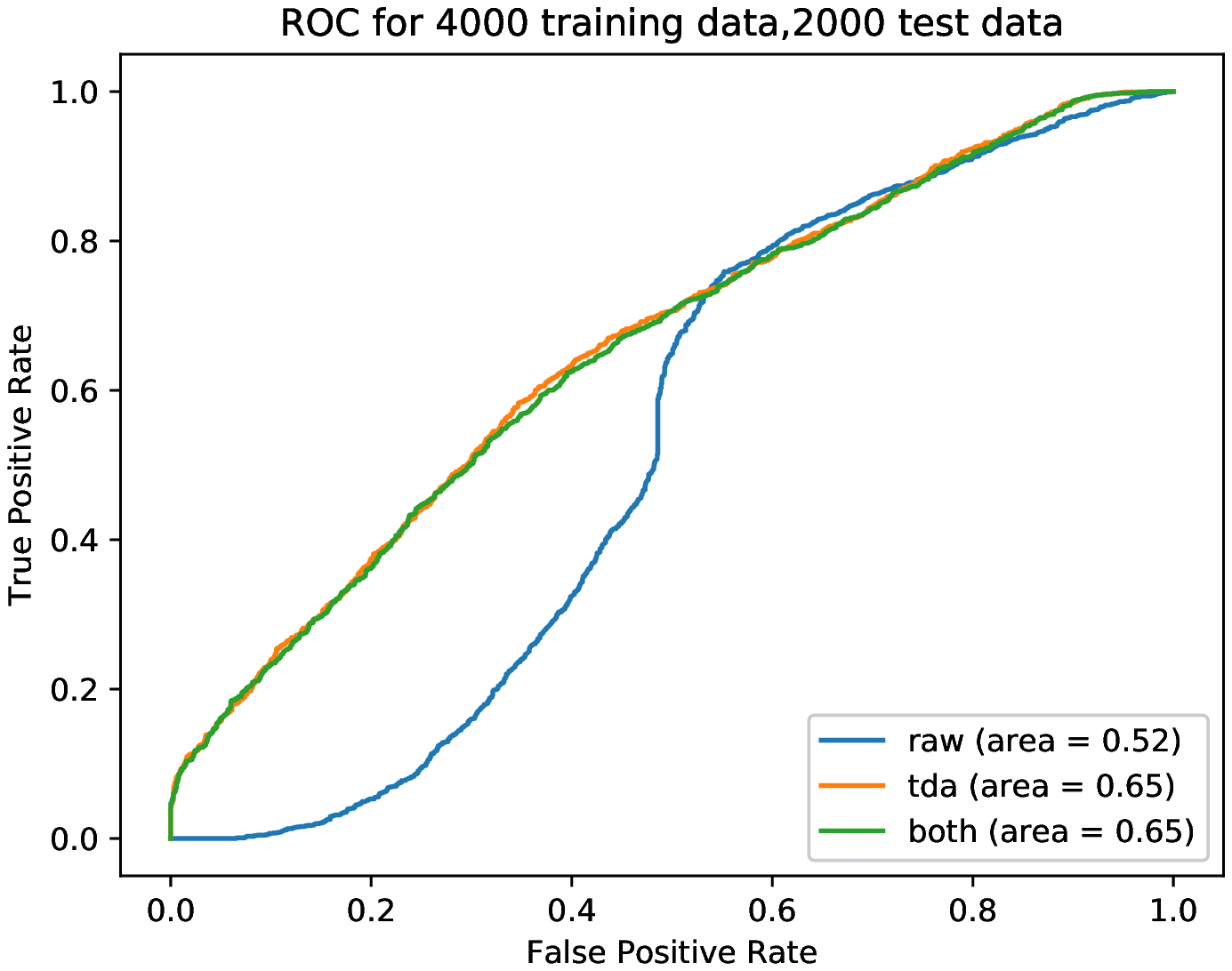}{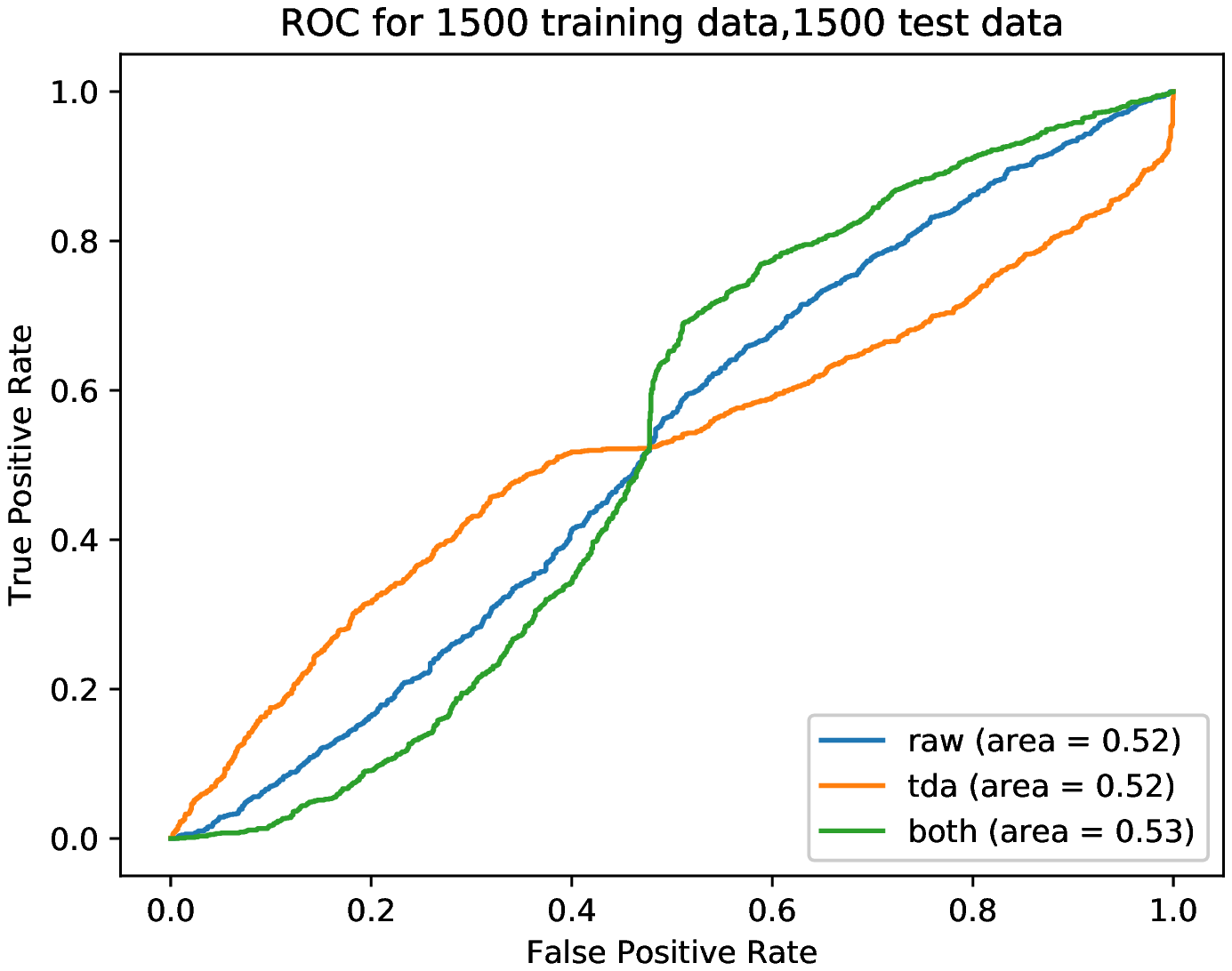}{0.32}{SNR of $6.874$. }{fig:roc25}

\threefigl[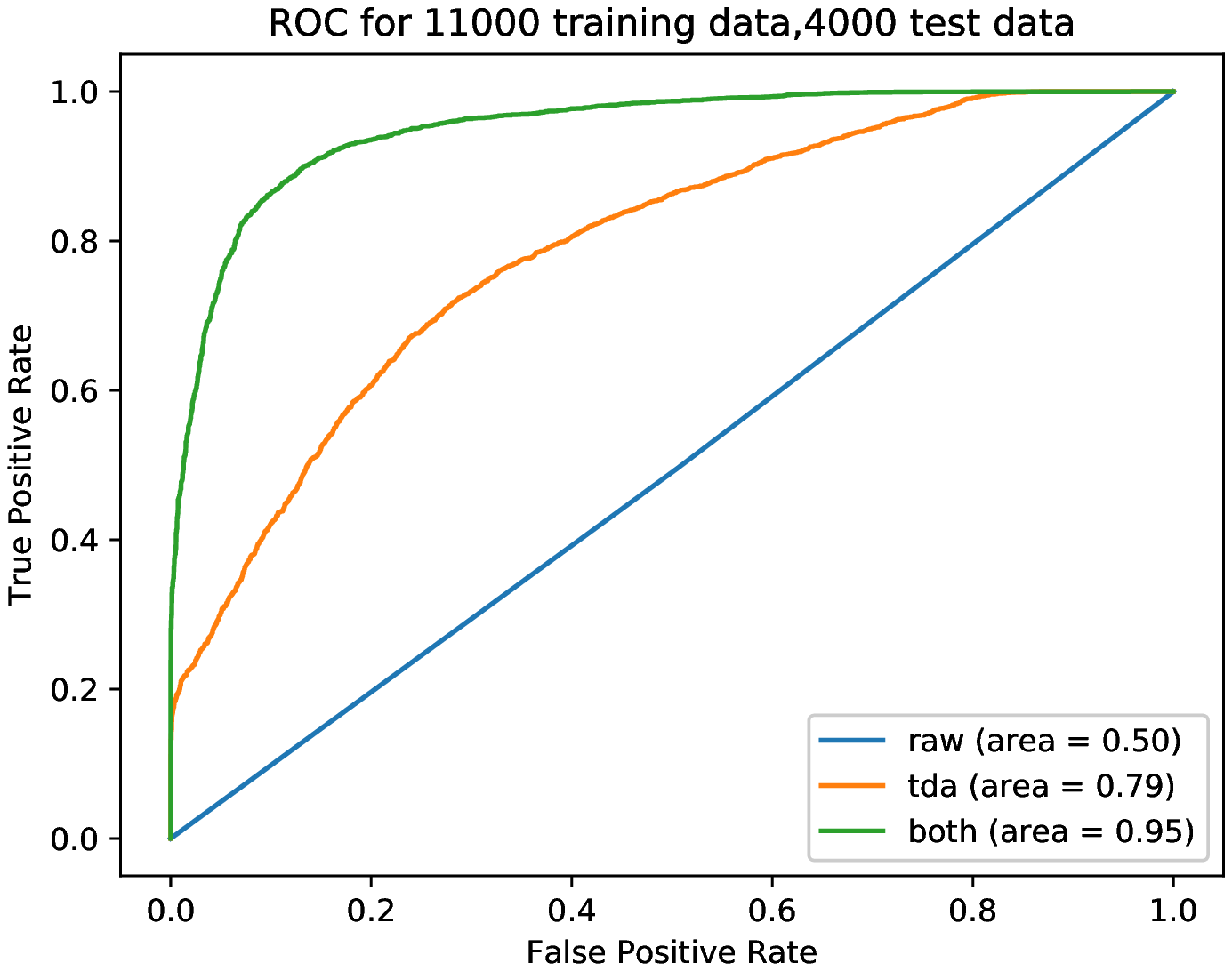]{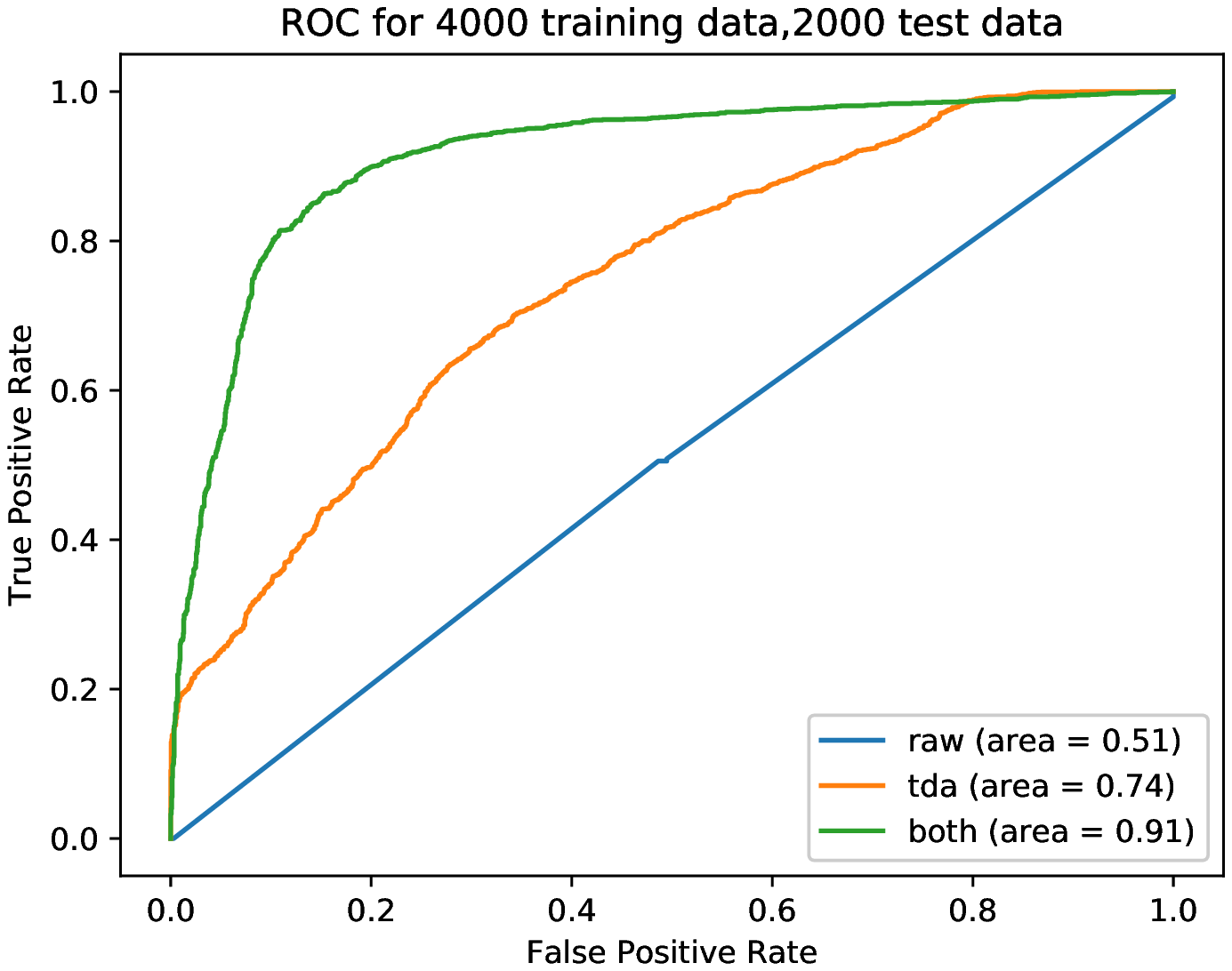}{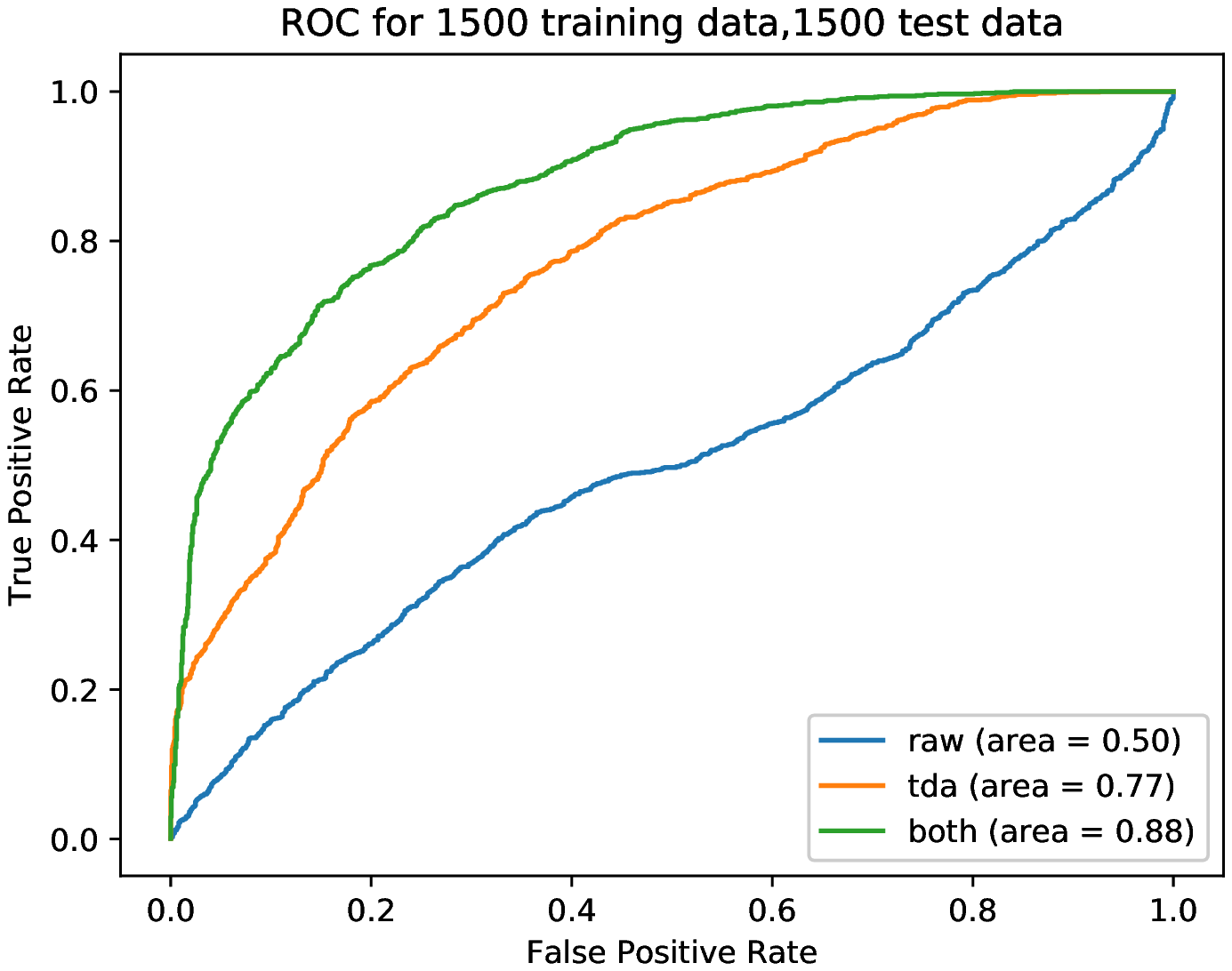}{0.32}{ SNR of $9.782$.}{fig:roc35}

%\threefig[figs/roc0.4ndat15000.eps]{figs/roc0.4ndat6000.eps}{figs/roc0.4ndat3000.eps}{0.32}{Signal level at 0.4 $1500$ mass ratios}

\threefigl[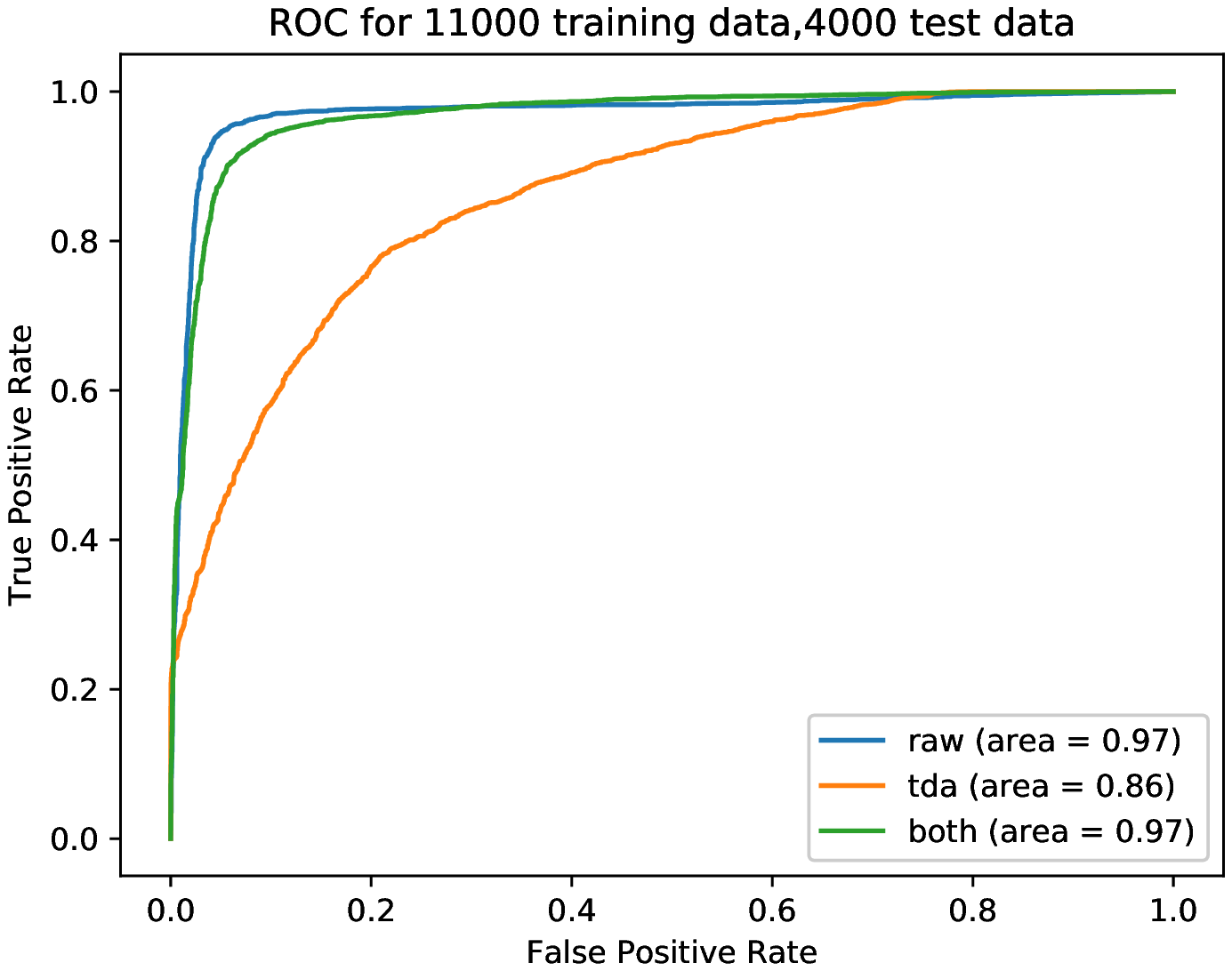]{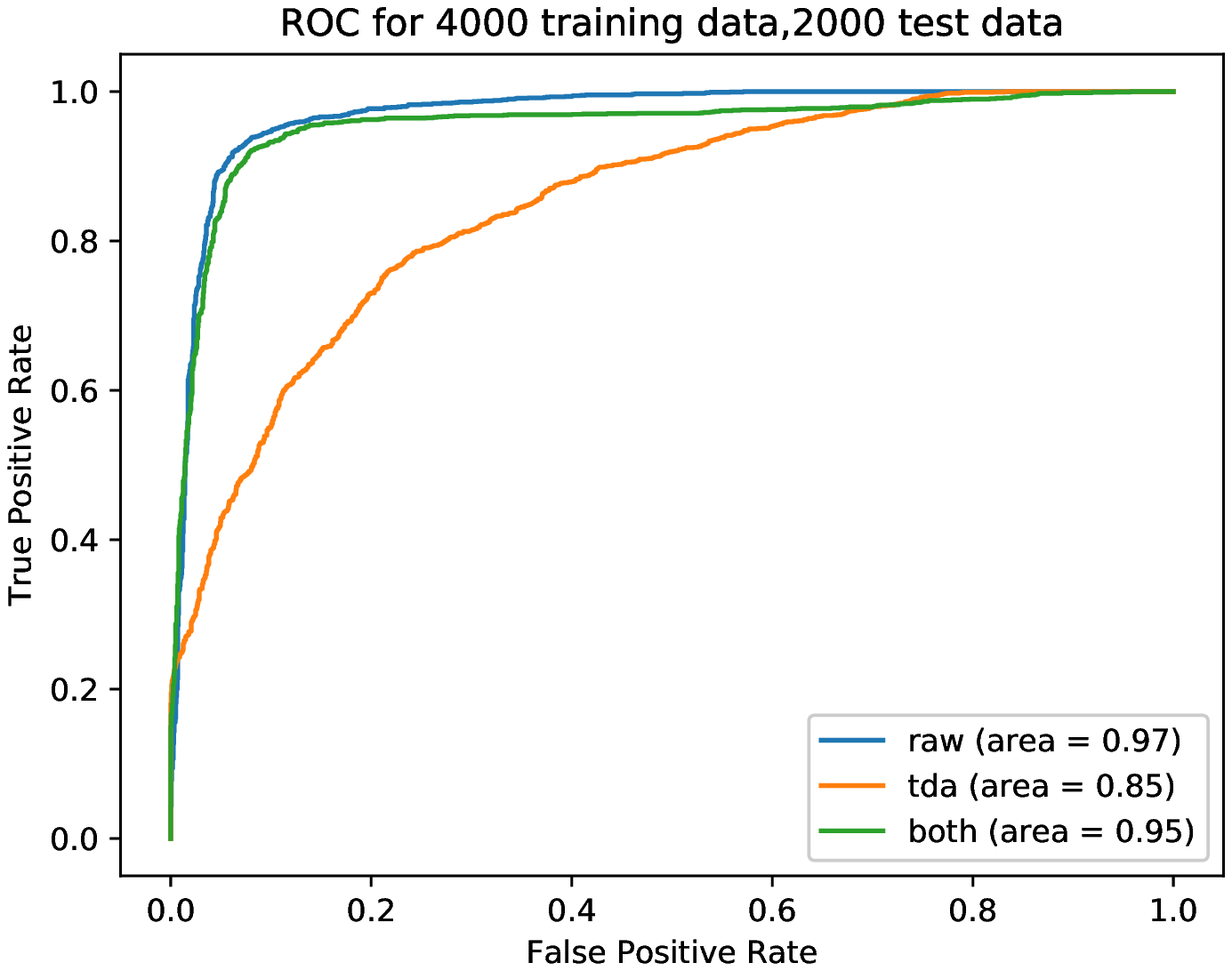}{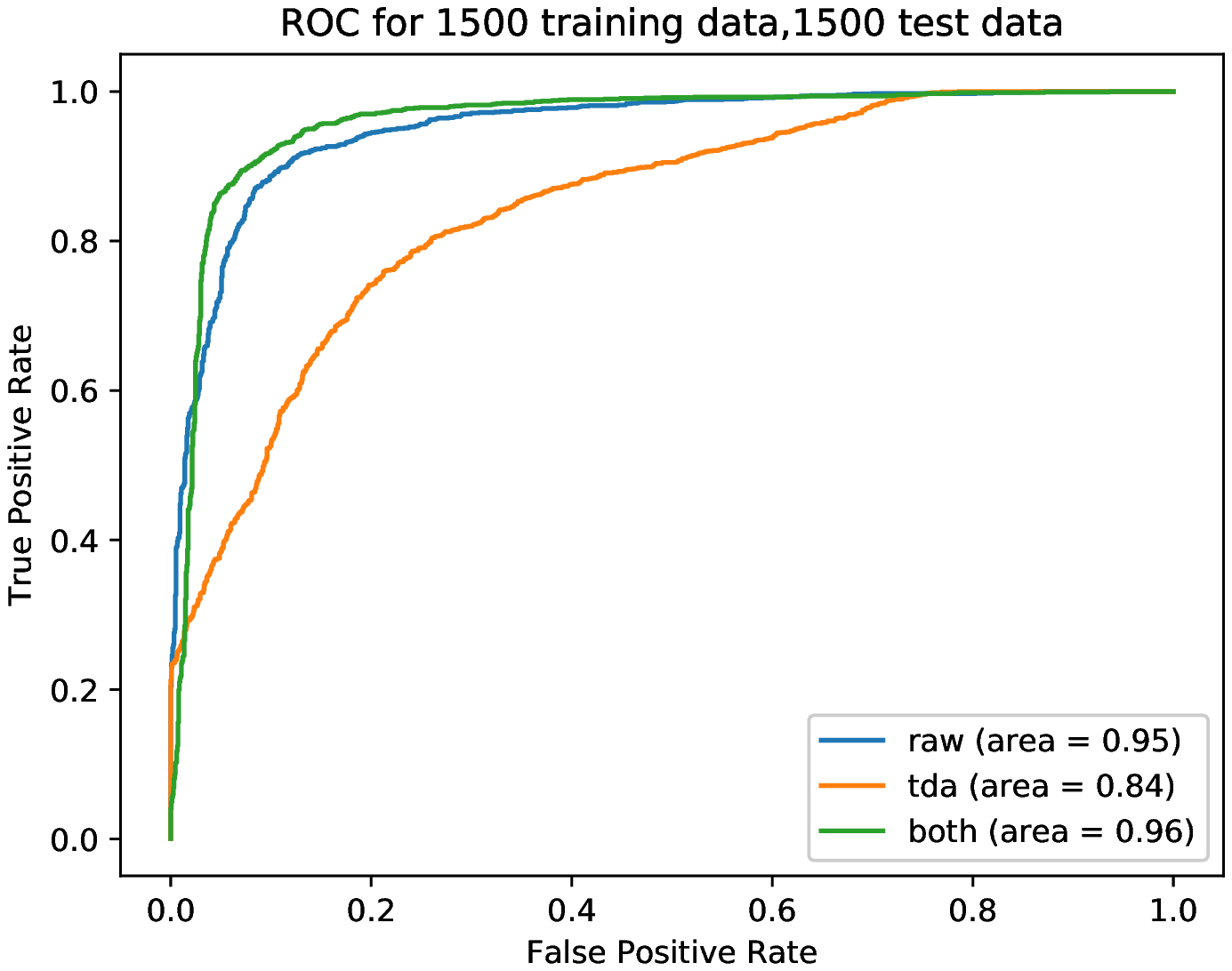}{0.32}{SNR of $12.519$.}{fig:roc45}

As these examples indicate, the proposed method yields a significant improvement over the original CNN method.  It could be very useful for pre-screening the interferometer data-streams to locate potentially interesting windows before more costly analysis. It is not limited to the detection of black-hole mergers, as many interesting astrophysical sources of gravitational wave produce a chirp type of signature \cite{ligoneutron}.  

%%%%% HERE HERE HERE 

%\threefig[figs/roc0.4ndat15000.eps]{figs/roc0.4ndat6000.eps}{figs/roc0.4ndat3000.eps}{0.32}{Noise level at 0.4}

%%%%% HERE HERE HERE 
%\threefig[figs/roc0.45ndat15000.eps]{figs/roc0.45ndat6000.eps}{figs/roc0.45ndat3000.eps}{0.32}{Noise level at 0.45}

\begin{acknowledgments}
The authors thank Scott Field for providing data and for the helpful discussion. This work was supported by National Research Foundation.  %also dev of the gw tools
\end{acknowledgments}

%\appendix

%\section{Appendixes}

%\begin{equation}

%\end{equation}

%\begin{subequations}
%\begin{eqnarray}

%\end{eqnarray}
%\end{subequations}

%\nocite{*}
\bibliography{gw}% Produces the bibliography via BibTeX.

%\bibliography{gw}{}
%\bibliographystyle{plain}

\end{document}